\newcommand{\figwsingle}{0.7}
\newcommand{\figwdouble}{0.5}
\newcommand{\figwf}{0.25}
\documentclass[preprint,3p,times,]{elsarticle}

\usepackage{amssymb}
\usepackage{xcolor}
\usepackage{subfig}
\usepackage{blindtext}
\usepackage{import}
\usepackage{pdfpages}
\usepackage{float}
 \usepackage[labelfont=bf]{caption}
 \usepackage{amsmath}
 \usepackage[autostyle]{csquotes}
 \definecolor{rev2}{RGB}{21,119,40}
 \DeclareUnicodeCharacter{2009}{\,} 
 \DeclareUnicodeCharacter{2212}{-}

\usepackage{natbib}
\usepackage{draftwatermark}

\SetWatermarkText{Preprint}
\SetWatermarkScale{1}
\setcitestyle{authoryear,open={(},close={)}} 

\journal{..}

\begin{document}

\begin{frontmatter}



\title{\textcolor{black}{On the surface chemisorption of oxidizing fine iron particles: insights gained from molecular dynamics simulations}}





\author[First]{L.C. Thijs\corref{cor1}}

\author[Second]{E. Kritikos}

\author[Second]{A. Giusti}

\author[First]{W.J.S Ramaekers}

\author[First]{J.A. van Oijen}

\author[First]{L.P.H de Goey}

\author[First]{X.C. Mi}

\cortext[cor1]{Corresponding authors.\\l.c.thijs@tue.nl (L.C. Thijs)}
\address[First]{Department of Mechanical Engineering, Eindhoven University of Technology, P.O. Box 513, NL-5600 MB, Eindhoven, Netherlands}

\address[Second]{Department of Mechanical Engineering, Imperial College London, London SW7 2AZ, United Kingdom}
\begin{abstract}
Molecular dynamics (MD) simulations are performed to investigate the thermal and mass accommodation coefficients (TAC and MAC, respectively) for the combination of iron(-oxide) and air. The obtained values of TAC and MAC are then used in a point-particle Knudsen model to investigate the effect of \textcolor{black}{chemisorption and the Knudsen transition regime }on the combustion behavior of (fine) iron particles. The thermal accommodation for the interactions of $\mathrm{Fe}$ with $\mathrm{N_2}$ and $\mathrm{Fe_xO_y}$ with $\mathrm{O_2}$ is investigated for different surface temperatures, while the mass accommodation coefficient for iron(-oxide) with oxygen is investigated for different initial oxidation stages $Z_\mathrm{O}$, which represents the molar ratio of $\mathrm{O}/\left(\mathrm{O} + \mathrm{Fe}\right)$, and different surface temperatures. The MAC decreases \textcolor{black}{fast from unity to 0.03 as $Z_\mathrm{O}$ increases from 0 to 0.5 and then diminishes as $Z_\mathrm{O}$ further increases to 0.57}. By incorporating the MD-informed accommodation coefficients into the single iron particle combustion model, the oxidation beyond $Z_\mathrm{O} = 0.5$ (from stoichiometric $\mathrm{FeO}$ to $\mathrm{Fe_3O_4}$) is modeled. A new temperature evolution for single iron particles is observed compared to results obtained with previously developed continuum models. Specifically, results of the present simulations show that the oxidation process continues after the particle reaching the peak temperature, while previous models predicting that the maximum temperature was attained when the particle is oxidized to $Z_\mathrm{O} = 0.5$. Since the \textcolor{black}{rate of oxidation} slows down as the MAC decreases with an increasing oxidation stage, the rate of heat loss exceeds the rate of heat release upon reaching the maximum temperature, \textcolor{black}{while the particle is not yet oxidized to $Z_\mathrm{O} = 0.5$}. Finally, the effect of transition-regime heat and mass transfer on the combustion behavior of fine iron particles is investigated and discussed.
\end{abstract}

\begin{keyword}
Iron particle; Knudsen transition heat and mass transfer; Thermal accommodation coefficient; Mass accommodation coefficient; Molecular dynamics; Metal combustion; Metal-enabled cycle of renewable energy
\end{keyword}

\end{frontmatter}


\section{Introduction}
Iron powder is considered as a promising metal fuel since it is inherently carbon-free, recyclable, compact, cheap and widely available \citep{Bergthorson2015}. To design and improve real-world iron-fuel burners, an in-depth understanding of the fundamentals underlying the combustion of fine iron particles is required.

Over the past five years, the interest in using iron powder as a circular carrier of renewable energy has drastically increased. \cite{Soo2017}, \cite{Tang2009}, \cite{McRae2019}, \cite{Toth2020} and \cite{Li2020} performed experiments with iron dust flames to study fundamental characteristics of iron combustion, such as flame structure and flame propagation. \cite{Toth2020} and \cite{Li2020} identified the formation of nano-particles in their experiments, where they observed halos of nano-particles surrounding the burning iron particles. To gain a more in-depth understanding of the oxidation processes of iron particles, the canonical configuration of single iron particle combustion has been investigated by multiple researchers. \cite{Ning2020, Ning2021} performed single particle combustion experiments where the particles are ignited by a laser. They showed that the duration of burning process is sensitive to the surrounding oxygen concentration. They also observed that the maximum particle temperature increases with gas-phase oxygen concentration while reaching a plateau at sufficiently elevated oxygen concentrations. Formation of oxide nano-particles during the combustion was also recorded. \cite{Li2022} investigated the ignition and combustion process of single micron-sized iron particles in the hot gas flow of a burned methane-oxygen-nitrogen mixture, and drew similar conclusions as \cite{Ning2020}. \cite{Panahi2022} used a drop-tube furnace to burn iron particles at a high gas temperature ($\approx 1350~\mathrm{K}$) with oxygen concentrations of $21\%$, $50\%$ and $100\%$. They showed that particle temperature increases significantly when the oxygen concentration increases from $21\%$ to $50\%$, but barely further increases when increasing from $50\%$ to $100\%$.

In the past few years, the number of theoretical models for single iron particles has increased. \cite{Mi2022} investigated the ignition behavior of iron particles via solid-phase oxidation kinetics described by a parabolic rate. \cite{Philyppe2022} extended this model and investigated the ignition behavior of fine iron particles in the Knudsen transition regime. They stated that the transition effect on the ignition characteristics becomes important if the particle diameter is below $30$~\textmu m. \cite{Senyurt2022} studied the ignition of metal particles other than iron in the transition regime, and stated that transition effects could always be neglected for very large (i.e., $>200$\textmu m) particles. While these models only focused on the ignition, \cite{Soo2018} developed a generic model for the full combustion behavior for non-volatile solid-fuel particles, wherein the particle oxidation rate depends on reaction kinetics at the surface of the particle and the external diffusion of oxygen to the particle. \cite{Hazenberg2020} further extended this model by taking into account the growth of the particle during oxidation. In \cite{ThijsPCI_2022}, a boundary layer resolved model was developed so that mass and heat transfer are accurately modeled, including Stefan flow. Temperature-dependent properties were used for the gas- and condensed-phase species and evaporation was implemented to investigate the formation of nano-sized iron-oxides products. It was shown that, although the particle temperature remains below the boiling point of iron(-oxide), a non-negligible amount of iron mass is lost due to the evaporation of the particle. Furthermore, it was shown that when only the conversion of $\mathrm{Fe}$ to $\mathrm{FeO}$ is considered up to the maximum temperature, and when internal transport is neglected, a good agreement was obtained with the experimental data of \citep{Ning2021} for the combustion time and maximum temperature as a function of oxygen concentration. The further oxidation after the particle peak temperature was, however, not modeled. In a later work of \citep{Thijs2022}, the point-particle model of \cite{Hazenberg2020} was extended, with aid of a boundary-layer-resolved model, to include temperature-dependent properties, slip velocity, Stefan flow, and evaporation. \textcolor{black}{\cite{Gool2022} extended the latter particle model and modeled the particle composition as being uniform and in thermodynamic equilibrium, enabling the oxidation beyond stoichiometric $\mathrm{FeO}$.}

\textcolor{black}{For iron particle combustion, it has been hypothesized that the oxidation rate of an iron droplet is the result of an interplay among three mechanisms: (1) External diffusion of $\mathrm{O_2}$ from the ambient gas to particle surface, (2) surface chemisorption of $\mathrm{O_2}$, and (3) internal transport of Fe and O atoms \citep{Sun2000}. In the previously discussed numerical models, it assumed that mechanism (1) is the rate limiting step, and that mechanisms (2) and (3) are infinitely fast. However, these assumptions are not well investigated.} Furthermore, in most of the previously discussed models for iron-particle combustion, the continuum assumption is used to describe the transport processes. It is known that, when the size of the particle becomes too small, modeling the heat and mass transfer using the continuum approach becomes invalid. The particle radius, $r_\mathrm{p}$, compared to the mean free path length of the gas molecules, $\lambda_\mathrm{MFP}$, is described by the Knudsen number \citep{Liu2006}
\begin{equation}
    \mathrm{Kn} = \lambda_\mathrm{MFP}/r_\mathrm{p}.
\end{equation}
Typically, when $\mathrm{Kn}$ is larger than 0.01 \citep{Liu2006}, the continuum approach is invalid. An elaborate review of modeling heat transfer in the transition regime for nano-particles in the context of laser-induced incandescence is presented by \cite{Liu2006}. In the previous study of \cite{ThijsPCI_2022}, only relatively large particles of $40$ and $50$\textmu m were considered, ensuring that the transition-regime heat and mass transfer has a negligible effect. While \cite{Philyppe2022} and \cite{Senyurt2022} studied the transition effect on the ignition behavior of metal particles, such a study was not performed for the complete combustion process of single iron particles. 

The heat and mass transfer in the free-molecular regime is dependent on the thermal and mass accommodation coefficients. The average energy transfer when gas molecules scatter from the surface is described by the thermal accommodation coefficient (TAC). The mass accommodation coefficient (MAC) or absorption coefficient is defined as the fraction of incoming oxygen molecules that are absorbed (accommodated) rather than reflected when they collide with the iron surface. Molecular dynamics simulation can be used to investigate these accommodation coefficients. \textcolor{black}{Multiple studies have been performed with molecular dynamics simulations to investigate the metal surface-gas interactions. In the case of highly-rarefied gases ($\mathrm{Kn} > 10$), the interaction between one gas molecule and the surface is considered. \cite{Chirita1993} performed one of the pioneering studies that consider molecular beam simulations to simulate incident low energy argon molecules on a nickel surface. Later, multiple authors \citep{Daun2009, Daun2012, Sipkens2014, Sipkens2018, Thoudam2020, Peddakotla2019, Mane2018} investigated the gas-surface interactions, considering one gas molecules issuing from a Maxwell-Boltzmann distribution.} \cite{Daun2012} showed that the TAC obtained with such a molecular dynamics simulation well agrees with experimental data. \textcolor{black}{In less rarefied gases ($\mathrm{Kn} < 1$) gas-gas interactions become important in addition to the surface-gas interactions, and can impact the accommodation coefficients \citep{Nejad20202, Sun2008, Finger2006}}. Alas, to the authors' knowledge, a systematic analysis of the TAC and MAC for a system of iron(-oxide) surface exposed to air has not yet been reported. Such a study of the MAC is also of importance to derive effective chemical kinetics governing the rate of further oxidation beyond the stoichiometry of $\mathrm{FeO}$, which are missing in literature.

\textcolor{black}{To study the effect of chemisorption and the effect of the free-molecular regime}, molecular dynamics simulations are performed to determine the TAC and MAC for an iron(-oxide)-air system. \textcolor{black}{Then, these values are used in a two-layer point-particle model to  examine the effect of chemisorption and the effect of the Knudsen transition regime on the combustion behavior of single iron particles. The paper is organized as follows. Section 2 describes the methodology of the two-layer model used to describe the combustion of single iron particles. In Section 3, the procedure for the molecular dynamics simulations used to determine the thermal and mass accommodation coefficients is discussed. Section 4 presents the results of the TAC and MAC obtained from these molecular dynamics simulations. In Section 5, the results of the single iron particle combustion model are discussed. Conclusions are provided in Section 6. }

\section{Model formulation for single iron particle combustion}
\begin{figure}[h]
    \centering
    {\includegraphics[width=0.5\columnwidth]{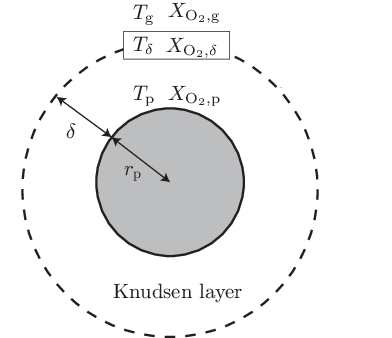}}
    \caption{The configuration for heat and mass transfer analysis considered in the Knudsen model.}%
     \label{fig:KnudsenConfig_Senyurt2022}
\end{figure}

In the current study, a two-layer point-particle model is used as shown in Figure \ref{fig:KnudsenConfig_Senyurt2022}. An iron particle oxidation model is coupled to a two-layer method to take into account the Knudsen transition regime. The iron particle oxidation model is based on the previous work \citep{Thijs2022}.

In \cite{Thijs2022}, only the oxidation from $\mathrm{Fe}$ into $\mathrm{FeO}$ was taken into account via
\begin{equation} \label{eq:Fe_into_FeO}
    \mathrm{R1}: \mathrm{Fe} + \frac{1}{2}\mathrm{O_2} \rightarrow  \mathrm{FeO}.
\end{equation}
Here, the oxidation mechanism is extended by considering the further oxidation into $\mathrm{Fe_3O_4}$. Once all the $\mathrm{Fe}$ is converted into $\mathrm{FeO}$, the $\mathrm{FeO}$ continues to oxidize into $\mathrm{Fe_3O_4}$ via 
\begin{equation}\label{eq:FeO_into_Fe3O4}
    \mathrm{R2}: 3\mathrm{FeO} + \frac{1}{2}\mathrm{O_2} \rightarrow  \mathrm{Fe_3O_4}.
\end{equation}
\textcolor{black}{The oxidation into $\mathrm{Fe_2O_3}$ is not considered since the re-solidification process is rather complex. Furthermore, the impact of the oxidation from $\mathrm{Fe_3O_4}$ into $\mathrm{Fe_2O_3}$ is rather small since only $10$\% of the total energy is released in this step.}
Based on the thermodynamics of the Fe-O system \citep{Wriedt1991}, it is known that iron-oxide can be formed in multiple crystalline structures, i.e., $\mathrm{FeO}$ (wustite), $\mathrm{Fe_3O_4}$ (magnetite), and $\mathrm{Fe_2O_3}$ (hematite). In the liquid phase of an Fe-O mixture, there are no clear crystalline structures. This region on the Fe-O phase diagram consists mainly of two different liquids denoted by L1 (iron) and L2 (iron-oxide).
Therefore, considering these phases as ``liquid $\mathrm{FeO}$'' and ``liquid $\mathrm{Fe_3O_4}$'' is not in line with the phase diagram.

The thermodynamic data for the L1 and L2 mixture are, however, unknown. In this work, the thermodynamic data for the L1 and L2 mixtures are therefore approximated to be a linear combination of the liquid-phase data of $\mathrm{Fe}$, $\mathrm{FeO}$, and $\mathrm{Fe_3O_4}$. The total enthalpy of a liquid iron particle, $H_\mathrm{liq,tot}$, is determined via
\begin{equation} \label{eq:total_enth_liq}
H_\mathrm{liq,tot} = m_{\mathrm{Fe}} h_{\mathrm{Fe}} + m_{\mathrm{FeO}} h_{\mathrm{FeO}} + m_{\mathrm{Fe_3O_4}} h_{\mathrm{Fe_3O_4}},
\end{equation}
with, $m_{{i}}$ the mass of species $i$ in the particle and $h_{{i}}$ the mass specific enthalpy of species $i$ (calculated with the NASA polynomials). \textcolor{black}{The liquid-phase polynomials are extrapolated to the boiling point of $\mathrm{FeO}$ ($T_\mathrm{p} = 3396 \: \mathrm{K}$)}.

The phase diagram of the Fe-O system is dependent on the molar ratio of $\mathrm{O}/\left(\mathrm{O} + \mathrm{Fe}\right)$, which defines the oxidation stage of the (liquid) iron particle. Here, this ratio is denoted by $Z_\mathrm{O}$, which is the elemental mole fraction of oxygen in the particle, and is defined as
\begin{equation}
    Z_\mathrm{O} = \frac{m_{\mathrm{O,p}}/M_{\mathrm{O}}}{m_{\mathrm{Fe,p}}/M_{\mathrm{Fe}} + m_{\mathrm{O,p}}M_{\mathrm{Fe}}},
\end{equation}
where $m_{\mathrm{O,p}}$ is the mass of oxygen, $m_{\mathrm{Fe,p}}$ the total mass of iron in the particle, and $M_{\mathrm{O}}$ and $M_{\mathrm{Fe}}$ are the molar mass of oxygen and iron, respectively. In this work, $Z_\mathrm{O}$ denotes the oxidation stage of the iron particle, where $Z_\mathrm{O} = 0.5$ represents ``liquid FeO'' and $Z_\mathrm{O} = 0.57$ ``liquid $\mathrm{Fe_3O_4}$''.

To calculate the total enthalpy of the particle, the rates of change of $m_{\mathrm{Fe}}$, $m_{\mathrm{FeO}}$, and $m_{\mathrm{Fe_3O_4}}$ need to be tracked. Since the \textcolor{black}{equilibrium} vapor pressure calculated with the thermodynamic data of liquid $\mathrm{Fe_3O_4}$ is orders-of-magnitude lower than those of liquid $\mathrm{Fe}$ and $\mathrm{FeO}$ \citep{Ning2021}, evaporation of $\mathrm{Fe_3O_4}$ is negligible compared to the evaporation of $\mathrm{Fe}$ and $\mathrm{FeO}$. \textcolor{black}{In \cite{ThijsPCI_2022} it was discussed that the oxidation of the gaseous iron will further inhibit the diffusion of oxygen towards the particle. To include this effect in the current point-particle model, it is taken into account that $\mathrm{Fe(g)}$ and $\mathrm{FeO(g)}$ will further oxidize and condense into $\mathrm{Fe_2O_3}$. The oxygen consumed by the gaseous iron(-oxide) species will limit the total mass transfer rate of oxygen. The mass transfer rate of oxygen consumed by the gaseous iron(-oxide) is given as \citep{Fujinawa2023}}
\textcolor{black}{\begin{equation}
    \dot{m}_\mathrm{O_2,vap} = s_\mathrm{Fe_2O_3,Fe(g)} \frac{\mathrm{d}m_{\mathrm{evap,\mathrm{Fe}}}}{\mathrm{d}t} + s_\mathrm{Fe_2O_3,FeO(g)}\frac{\mathrm{d}m_{\mathrm{evap,\mathrm{FeO}}}}{\mathrm{d}t},
\end{equation}}
\noindent \textcolor{black}{with $s_\mathrm{Fe_2O_3,Fe(g)}$ and $s_\mathrm{Fe_2O_3,FeO(g)}$ the stoichiometric mass ratios for the oxidation of respectively $\mathrm{Fe(g)}$ and $\mathrm{FeO(g)}$ into $\mathrm{Fe_2O_3}$.} The rate of change of $m_{\mathrm{Fe}}$, $m_{\mathrm{FeO}}$, and $m_{\mathrm{Fe_3O_4}}$ are related to the rate of oxidation and the rate of evaporation via
\begin{equation}
    \frac{\mathrm{d}m_{\mathrm{Fe}}}{\mathrm{d}t} = -\frac{1}{s_\mathrm{Fe,1}} \left(\dot{m}_\mathrm{O_2,1} \textcolor{black}{- \dot{m}_\mathrm{O_2,vap}}\right) - \frac{\mathrm{d}m_{\mathrm{evap,\mathrm{Fe}}}}{\mathrm{d}t},
\end{equation}
\textcolor{black}{\begin{equation}
    \frac{\mathrm{d}m_{\mathrm{FeO}}}{\mathrm{d}t}
    \begin{cases}
    -\frac{1}{s_\mathrm{FeO,1}} \left(\dot{m}_\mathrm{O_2,1} \textcolor{black}{- \dot{m}_\mathrm{O_2,vap}}\right) - \frac{\mathrm{d}m_{\mathrm{evap,\mathrm{FeO}}}}{\mathrm{d}t}& \text{if } Z_\mathrm{O} < 0.5,\\
     -\frac{1}{s_\mathrm{FeO,2}} \left(\dot{m}_\mathrm{O_2,2} \textcolor{black}{- \dot{m}_\mathrm{O_2,vap}}\right) - \frac{\mathrm{d}m_{\mathrm{evap,\mathrm{FeO}}}}{\mathrm{d}t}& \text{if } Z_\mathrm{O} > 0.5,
\end{cases}
\end{equation}}

\begin{equation}
    \frac{\mathrm{d}m_{\mathrm{Fe_3O_4}}}{\mathrm{d}t} = -\frac{1}{s_\mathrm{Fe_3O_4,2}} \left(\dot{m}_\mathrm{O_2,2} \textcolor{black}{- \dot{m}_\mathrm{O_2,vap}}\right),
\end{equation}
with $\dot{m}_{\mathrm{O_2},j}$ the mass transfer rate of oxygen to the particle via either reaction R1 (denoted by 1) or R2 (denoted by 2) and $s_{i,j}$ is the stoichiometric mass ratio of species $i$ for reaction $j$. The total mass transfer rate of oxygen is given by $\dot{m}_\mathrm{O_2} = \dot{m}_\mathrm{O_2,1} + \dot{m}_\mathrm{O_2,2}$. Note that R1 and R2 are sequential reactions in this work, \textcolor{black}{where R2 starts if $Z_\mathrm{O} > 0.5$. Therefore, either $\dot{m}_\mathrm{O_2,1}$ or $\dot{m}_\mathrm{O_2,2}$ equals zero.}

To calculate $Z_\mathrm{O}$, the mass of $\mathrm{O}$ and $\mathrm{Fe}$ in the particle must be known. The rate of change of $\mathrm{O}$ is related to the rate of oxidation and the rate of evaporation of $\mathrm{FeO}$, while the rate of change of $m_{\mathrm{Fe}}$ which is in the particle, is only affected by evaporation
\begin{equation}
    \frac{\mathrm{d}m_{\mathrm{O,p}}}{\mathrm{d}t} = \dot{m}_\mathrm{O_2} - \frac{M_\mathrm{O}}{M_\mathrm{FeO}}\frac{\mathrm{d}m_{\mathrm{evap,\mathrm{FeO}}}}{\mathrm{d}t},
\end{equation}
\begin{equation}
    \frac{\mathrm{d}m_{\mathrm{Fe,p}}}{\mathrm{d}t} = - \frac{\mathrm{d}m_{\mathrm{evap,\mathrm{Fe}}}}{\mathrm{d}t} -  \frac{M_\mathrm{Fe}}{M_\mathrm{FeO}}\frac{\mathrm{d}m_{\mathrm{evap,\mathrm{FeO}}}}{\mathrm{d}t},
\end{equation}
with $M_\mathrm{FeO}$ the molar mass of $\mathrm{FeO}$. Temperature-dependent density functions are used to relate the mass of the particle to the volume and diameter, see \cite{Gool2022} for the specific polynomials.

The rate of change of the particle enthalpy is described by
\begin{equation}
    \frac{\mathrm{d}H_{\mathrm{p}}}{\mathrm{d}t} = q + q_\mathrm{rad} + \dot{m}_\mathrm{O_2} h_\mathrm{O_2} - \sum_i h_{i,v} \frac{\mathrm{d}m_{\mathrm{evap}, i}}{\mathrm{d}t}, 
\end{equation}
with $q$ the heat transfer rate, $q_\mathrm{rad}$ the radiative heat transfer rate, $h_\mathrm{O_2}$ the mass-specific enthalpy of the consumed oxygen and $h_{i,v}$ the mass-specific enthalpy of the evaporated species. 

To model the mass transfer rate $\dot{m}_\mathrm{O_2}$ and heat transfer rate $q$, a two-layer method is used. Figure \ref{fig:KnudsenConfig_Senyurt2022} illustrates the configuration which is used for the two-layer model. The iron particle is surrounded by a spherical Knudsen layer $\delta$ with a thickness equal to the mean free path $\lambda_\mathrm{MFP}$ of the gas molecules \citep{Liu2006}
\begin{equation}
    \lambda_\mathrm{MFP} = \frac{k_\delta}{p} \frac{\gamma_\delta - 1}{9 \gamma_\delta - 5}\sqrt{\frac{8 \pi m_\mathrm{O_2} T_\delta}{k_b}},
\end{equation}
with $k_\delta$ and $\gamma_\delta$ being the thermal conductivity and specific heat ratio derived at the Knudsen layer, $p$ the ambient pressure, $m_\mathrm{O_2}$ the mass of an oxygen molecule, and $k_b$ the Boltzmann constant.

At the Knudsen layer, the gas has a temperature $T_\delta$ and mole fraction $X_\delta$. In the two-layer method, it is assumed that the heat and mass transfer across the Knudsen layer is quasi-steady \citep{Liu2006}. Therefore, the mass balance at $r = \delta$ yields $\dot{m}_\mathrm{O_2,FM} = \dot{m}_\mathrm{O_2,c}$, with $\dot{m}_\mathrm{O_2,FM}$ the mass transfer rate in the free-molecular regime and $\dot{m}_\mathrm{O_2,c}$ the mass transfer rate in the continuum regime. Similarly, the heat balance can be written as $q_\mathrm{FM} = q_\mathrm{c}$.

The mass transfer rate and heat transfer rate in the continuum regime can be described as \citep{Thijs2022,Senyurt2022}
\begin{equation} \label{eq:mdot_cont}
    \dot{m}_\mathrm{O_2,c} = 2 \pi \mathrm{Sh} \left(\delta + r_p\right) \rho_\mathrm{O2,f} D_\mathrm{f} \left(X_\mathrm{O_2,g} - X_{\mathrm{O_2},\delta}\right),
\end{equation}
\begin{equation}\label{eq:qdot_cont}
    q_\mathrm{c} = 2 \pi \mathrm{Nu} \left(\delta + r_p\right) k_\mathrm{f} \left(T_\mathrm{g} - T_\delta \right),
\end{equation}
with $T_\mathrm{g}$ is the gas temperature, $X_\mathrm{O_2,g}$ the mole fraction of oxygen in the gas phase,  $\mathrm{Nu}$ and $\mathrm{Sh}$ the Nusselt and Sherwood numbers, respectively and  $\rho_\mathrm{O2,f}$, $D_\mathrm{f}$ and $k_\mathrm{f}$ the density of oxygen, the mixture-averaged diffusion coefficient and the thermal conductivity derived at the film temperature and composition. \textcolor{black}{To include the effect of a slip velocity and the oxygen consumption induced Stefan flow towards the particle, the Nusselt and Sherwood numbers are corrected by means of the Reynolds number and the mass and heat Spalding numbers. The Stefan flow corrected Nusselt ans Sherwood numbers are determined as
\begin{equation}
    \mathrm{Nu} = \mathrm{Nu^*} \ln\left(1 + B_\mathrm{T}\right)/B_\mathrm{T},
\end{equation}
\begin{equation}
    \mathrm{Sh} = \mathrm{Sh^*} \ln\left(1 + B_\mathrm{m}\right)/B_\mathrm{m},
\end{equation}
with $\mathrm{Nu^*}$ and $\mathrm{Sh^*}$ being Reynolds number dependent correlations as described in \cite{Thijs2022}. The mass Spalding number is defined as \citep{bird2002}
\begin{equation}
    B_\mathrm{m} = \frac{Y_\mathrm{O_2,g} - Y_\mathrm{O_2,p}}{Y_\mathrm{O_2,p} - 1}.
\end{equation}
Analogous to the mass Spalding number, a Spalding heat transfer number can be derived as \citep{bird2002}
\begin{equation}
    1 + B_\mathrm{T} = \left(1 + B_\mathrm{m}\right)^{\phi_m},
\end{equation}
where $\phi_m = \frac{c_\mathrm{p,ox}}{c_\mathrm{p}}\frac{1}{\mathrm{Le}}\frac{\mathrm{Sh}}{\mathrm{Nu}}$ and $\mathrm{Le} = \frac{\mathrm{Sc}}{\mathrm{Pr}}$ is the Lewis number}. A constant slip velocity \textcolor{black}{in time} is used, \textcolor{black}{based on the terminal velocity of the iron particle at ambient conditions}. For more details of the continuum model, the reader is referred to \cite{Thijs2022}.

The mass transfer rate in the free-molecular regime can be described as \citep{Senyurt2022}
\begin{equation}
    \dot{m}_\mathrm{O_2,FM} = \alpha_\mathrm{m} \pi r_p^2 v_\mathrm{\delta} \rho_\mathrm{O2,\delta} X_{\mathrm{O_2},\delta},
\end{equation}
with $\alpha_\mathrm{m}$ the mass accommodation coefficient and $v_{\delta}$ the velocity of the gas molecules calculated as 
\begin{equation}
    v_\delta = \sqrt{\frac{8 k_b T_\delta}{\pi m_\mathrm{O_2}}}.
\end{equation}

Similarly, the heat transfer rate in the free-molecular regime yields \citep{Liu2006}
\begin{equation}
    q_\mathrm{FM} = \alpha_\mathrm{T} \pi r_\mathrm{p}^2 p \sqrt{\frac{k_b T_\delta}{8 \pi m_\mathrm{O_2}}} \frac{\gamma^* + 1}{\gamma^* - 1}\left(\frac{T_\mathrm{p}}{T_\delta} - 1 \right),
\end{equation}
with $\alpha_\mathrm{T}$ the thermal accommodation coefficient and $\gamma^*$ the averaged specific heat ratio \citep{Liu2006}. It is assumed that the slip velocity and Stefan flow do not affect the free-molecular transport and the effect of the transition regime on the evaporation rate is neglected.

\textcolor{black}{To account for the large variation of temperature and composition in the boundary layer, the transport properties are derived at film layer temperature ($T_\mathrm{f}$) and composition ($X_\mathrm{f}$), by using the 1/2 film averaging rule \citep{Thijs2022}. $T_\mathrm{f}$ and $X_\mathrm{f}$ are determined via
\begin{equation}
    T_\mathrm{f} = T_\mathrm{1} + 1/2\left(T_\mathrm{2} - T_\mathrm{1}\right),
\end{equation}
\begin{equation}
    X_\mathrm{f} = X_\mathrm{1} + 1/2\left(X_\mathrm{2} - X_\mathrm{1}\right).
\end{equation}
In the Knudsen layer, $T_\mathrm{1}$ and $X_\mathrm{1}$ are the values derived at the particle surface, while $T_\mathrm{2}$ and $X_\mathrm{2}$ are determined at the $\delta$ layer. For the continuum layer, $T_\mathrm{1}$ and $X_\mathrm{1}$ are the values derived at the $\delta$ layer, while $T_\mathrm{2}$ and $X_\mathrm{2}$ are determined at the far field. }

The two-layer method is implicit---a coupled system of nonlinear equations must be solved to find the mass and heat transfer rates. The molar fraction of oxygen $X_\mathrm{O_2,g}$ and the temperature $T_\mathrm{g}$ in the gas phase are known. Therefore, $X_{\mathrm{O_2},\delta}$ and $T_\delta$ remain unknown and should be found via an iterative method. In this work, the lsqnonlin method of MATLAB was used to solve the coupled system of nonlinear equations. The Cantera toolbox was used to calculate the transport properties in the gas phase.

\section{Model formulation for molecular dynamics simulations}\label{sec:method}
Molecular \textcolor{black}{dynamics simulations are performed to determine the TAC and MAC for the interactions between a single gas molecule and a iron(-oxide) surface. This is consistent with the free-molecular regime because gas-gas interactions are not considered.} \textcolor{black}{Because an $\mathrm{N_2}$ molecule has a strong triple bond, it is unlikely to be chemically absorbed by an iron(-oxide) surface.} \textcolor{black}{In the case of $\mathrm{NO_{x}}$ formation, different species like $\mathrm{N}$ and $\mathrm{NO}$ could possible interact with the iron particle \citep{Roy2009}.} In this study, only a TAC value for the $\mathrm{N_2}$ molecule in combination with an iron surface is determined, leaving the study of nitrogen oxides for future work. As a result, the $\mathrm{Fe}$-$\mathrm{N_2}$ interactions are modeled using non-reactive potentials. On the contrary, reactive molecular dynamics are considered for the $\mathrm{Fe_xO_y}$-$\mathrm{O_2}$ interactions to compute the TAC and MAC. \textcolor{black}{In reactive MD simulations, statistical and methodological errors can be introduced \citep{Kroger2017}. In this study, the thermal and mass accommodation coefficients were averaged over a number of instances in order to achieve small statistical uncertainties.} Large-scale Atomic/Molecular Massively Parallel Simulator (LAMMPS) \citep{LAMMPS} is used to perform the molecular dynamics simulations. 

\subsection{Thermal and mass accommodation coefficients}
As previously described, the free-molecular heat transfer rate is dependent on the TAC. The TAC describes the average energy transfer when gas molecules scatter from the surface and is defined as
\begin{equation}
    \alpha_\mathrm{T} = \frac{\left<E_0 - E_i\right>}{3k_\mathrm{B}\left(T_\mathrm{s} - T_\mathrm{g}\right)},
\end{equation}
with $\left<\cdot\right>$ denoting an ensemble average, $E_0$ the total energy of the scattered molecule, and $E_i$ the energy of the incident molecule. The denominator represents the maximum energy that could be transferred from the surface to the gas molecule, with $T_\mathrm{s}$ the surface temperature and $T_\mathrm{g}$ the gas temperature.


The free-molecular mass transfer rate is dependent on the MAC. The MAC or absorption coefficient is defined as the fraction of incoming oxygen molecules that, upon collision with the iron surface, are absorbed (accommodated) rather than reflected. The MAC is defined as
\begin{equation}
    \alpha_\mathrm{m} = \frac{n_\mathrm{abs,g}}{n_\mathrm{tot,g}},
\end{equation}
with $n_\mathrm{abs,g}$ the number of absorbed gas molecules and $n_\mathrm{tot,g}$ the total number of gas molecules colliding the surface. 

When iron is burned in air, both oxygen and nitrogen may contribute to the total TAC. \textcolor{black}{The thermal accommodation coefficients for $\mathrm{Fe}$ with $\mathrm{N_2}$ is independently determined from the interaction between $\mathrm{Fe_xO_y}$ with $\mathrm{O_2}$. The ratio between the numbers of collisions of oxygen and nitrogen with the surface is determined by the molar fraction, denoted by $X_\mathrm{O_2}$. In the case that an oxygen molecule is accommodated to the surface, it does not contribute to the thermal accommodation coefficient. To account for this effect, the total TAC is calculated as}
\begin{equation} \label{eq:TACTotal}
    \alpha_\mathrm{T,tot} = \left[1 - X_\mathrm{O_2}\left(1 - \alpha_\mathrm{m}\right)\right] \alpha_\mathrm{T,N_2} + X_\mathrm{O_2}\left(1 - \alpha_\mathrm{m}\right) \alpha_\mathrm{T,O_2}.
\end{equation}
\textcolor{black}{Note that if the oxygen molecule is chemically absorbed, the amount of enthalpy possessed by the absorbed oxygen is added to the total internal energy of the particle.}

\subsection{$\mathrm{Fe}$ - $\mathrm{N_2}$ interaction}
\begin{figure}[t]
    \centering
    {\includegraphics[width=0.35\columnwidth]{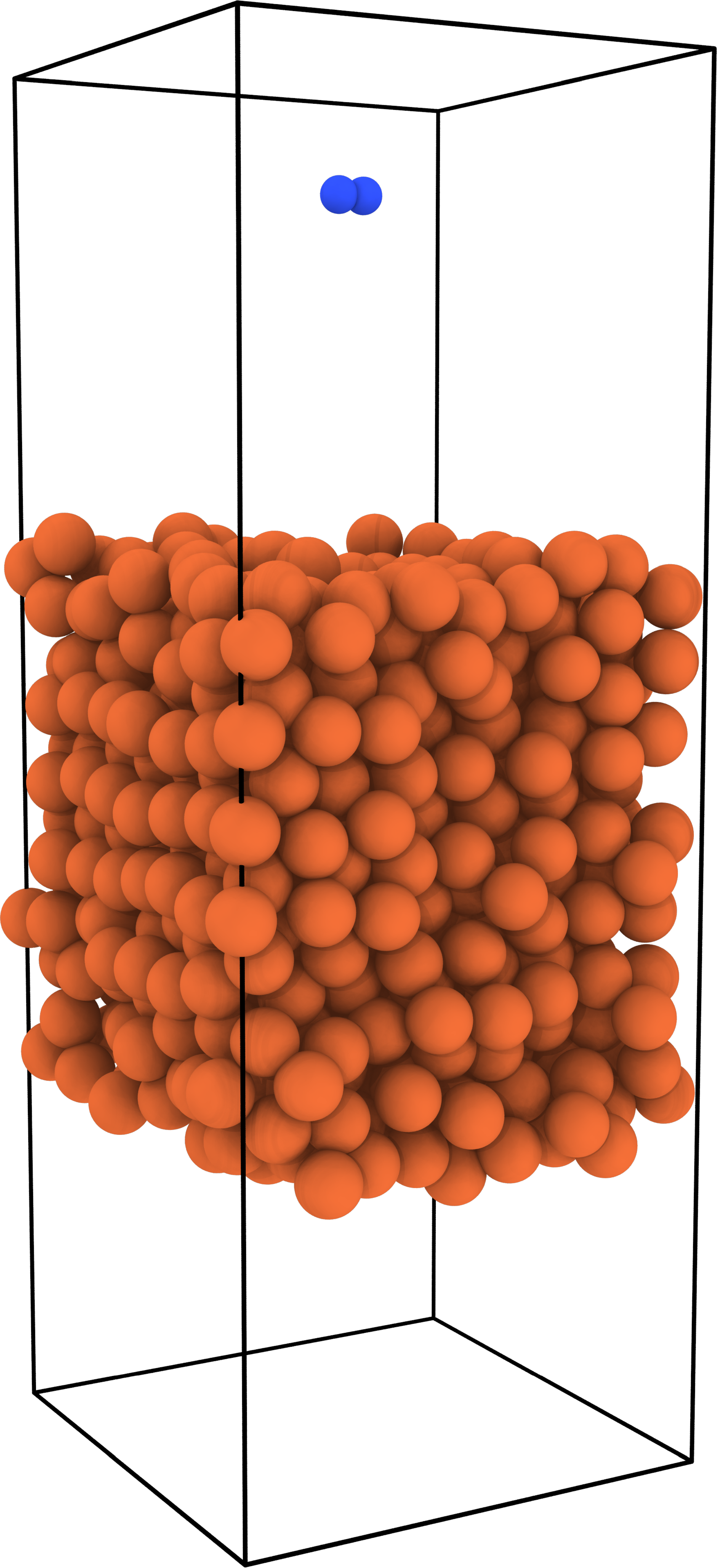}}
    \caption{Initial configuration showing a nitrogen molecule (blue particles) placed above an iron surface (brown particles), with $Z_\mathrm{O} = 0$ and $T_\mathrm{p} = 2000 \: \mathrm{K}$. The sizes of the particles equal the atomic radii. \textcolor{black}{CPK coloring is used to distinguish different chemical elements.}}%
     \label{fig:InitialConfig}
\end{figure}

To determine the thermal accommodation coefficients, the procedure as performed by \cite{Sipkens2018} is followed. Figure \ref{fig:InitialConfig} shows the initial configuration used to determine the thermal accommodation coefficient of the $\mathrm{Fe}$-$\mathrm{N_2}$ system. A molecular system is defined with a surface of 686 iron atoms initially arranged in a body-centered cubic (BCC) lattice, with a lattice constant of $2.856 \: \mathrm{\AA}$, and a nitrogen molecule, of which the latter is modeled as a rigid rotor. The gas molecule is positioned around $10\:\mathrm{\AA}$ above the surface, beyond the range of the potential well. In order to represent a specific surface temperature, a heating process is required for the iron surface to increase the kinetic energy of the system. \cite{Yan2017} showed that the phase change temperature obtained with MD simulations depends on the used heating process. Therefore, if the surface is expected to be in a solid phase, it is heated for 30 ps using the canonical (NVT) ensemble. To keep a constant temperature in the NVT ensemble simulations, the Nose-Hoover thermostat is applied on the translational degrees of freedom of the atoms with a temperature damping period of 100 fs. The warmed surfaces are then allowed to run first in the NVT ensemble and then in the micro-canonical (NVE) ensemble for 5~ps after which their state is saved to a file. If the surface is expected to be in a liquid phase, the surfaces are warmed using the Nose-Hoover thermostat for 30 ps to $2800\:\mathrm{K}$, equilibrated for 5~ps at $2800\:\mathrm{K}$ and then gradually cooled down to the target temperature within 10 ps. Six different surfaces, each with different initial velocities, are generated to obtain a statistically meaningful set of data. Then, incident gas molecules are introduced with their velocities sampled from the Maxwell-Boltzmann (MB) distribution. 500 cases per warmed surface are sampled, which result in 3000 data points per configuration. The equations of motion are advanced using the Verlet algorithm with a timestep of $1$~fs.

The interactions between the iron atoms are modeled using the embedded atom method (EAM) potential. The EAM force fields are based on the principle of embedding atoms within an electron cloud. The EAM potential reads
\begin{equation}
U_i = \frac{1}{2} \sum_{j\neq i} U_2\left(r_{ij}\right) - F\left(\sum_{j\neq i} \rho_j\left(r_{ij}\right) \right),
\end{equation}
where $U_2$ is a pairwise potential between atoms $i$ and $j$, $r_{ij}$ is the distance between atoms $i$ and $j$, $F$ is the embedding function, and $\rho_j$ is the contribution to the electron charge density from atom $j$. \cite{Sipkens2018} investigated the effect of different surface potentials on the characteristics of an iron lattice, and found that the EAM potential of \cite{Zhou2004} is the most robust choice since this potential well predicts the phase transitions and experimentally measured densities. Therefore, this work uses the EAM potential of \cite{Zhou2004} for the iron surface potentials.

\textcolor{black}{The gas-surface pairwise potentials are described by the Morse potential}
\begin{equation}
        U_{ij} = D_0 \left(1 - e^{-\alpha\left(r-r_0\right)} \right)^2,
\end{equation}
where $r_{0}$ is the equilibrium bond distance and $\alpha$ controls the width of the potential.

\cite{Daun2012} demonstrated the significance of using a well-defined gas-surface potential. They compared the TAC calculated with a Lennard-Jones 6-12 gas-surface potential with parameters derived from the Lorentz-Berthelot (LB) combination rules to the TAC calculated with a Morse potential with parameters derived from \textit{ab initio} technique. For the LJ potential with LB combination rules, the well depth $D_0$ and finite distance $\sigma$ were calculated by combining these values of the two individual atoms. For the \textit{ab initio} calculations a molecule was moved towards a surface and then the potential was calculated from first principles. \cite{Daun2012} showed that the TAC is overestimated when using a Lennard-Jones 6-12 gas-surface potential with parameters derived from the Lorentz-Berthelot (LB) combination rules, but found a good agreement with the values obtained from experiments when using a Morse potential with parameters derived from \textit{ab initio} technique. The \textit{ab initio} calculations for the $\mathrm{Fe}$ - $\mathrm{N_2}$ interaction were performed by \cite{Sipkens2014}. They found $D_0 = 2.162 \: \mathrm{meV}$, $\alpha = 0.932 \: \mathrm{\AA^{-1}}$ and $r_0 = 4.819\: \mathrm{\AA}$. 




\subsection{$\mathrm{Fe_xO_y}$ - $\mathrm{O_2}$ interaction} \label{sec:FexOy_O2_interaction}
Different iron-oxide surfaces will be generated to investigate the effect of the oxidation stage of the iron-oxide surface on the TAC and MAC. The surface oxidation stage is denoted by the elemental mole fraction of oxygen in the particle and can be calculated as
\begin{equation}
    Z_\mathrm{O} = \frac{n_{\mathrm{O,s}}}{n_\mathrm{tot,s}},
\end{equation}
with $n_\mathrm{O,s}$ being the number of oxygen atoms and $n_\mathrm{tot,s}$ the total number of atoms in the \textcolor{black}{complete domain. By means of this definition, $Z_\mathrm{O}$ remains constant in time if no new oxygen molecules are absorbed}.

\begin{figure*}
    \centering
        \subfloat[$0$ ps]
         {\includegraphics[width=\figwf\columnwidth , height=10cm]{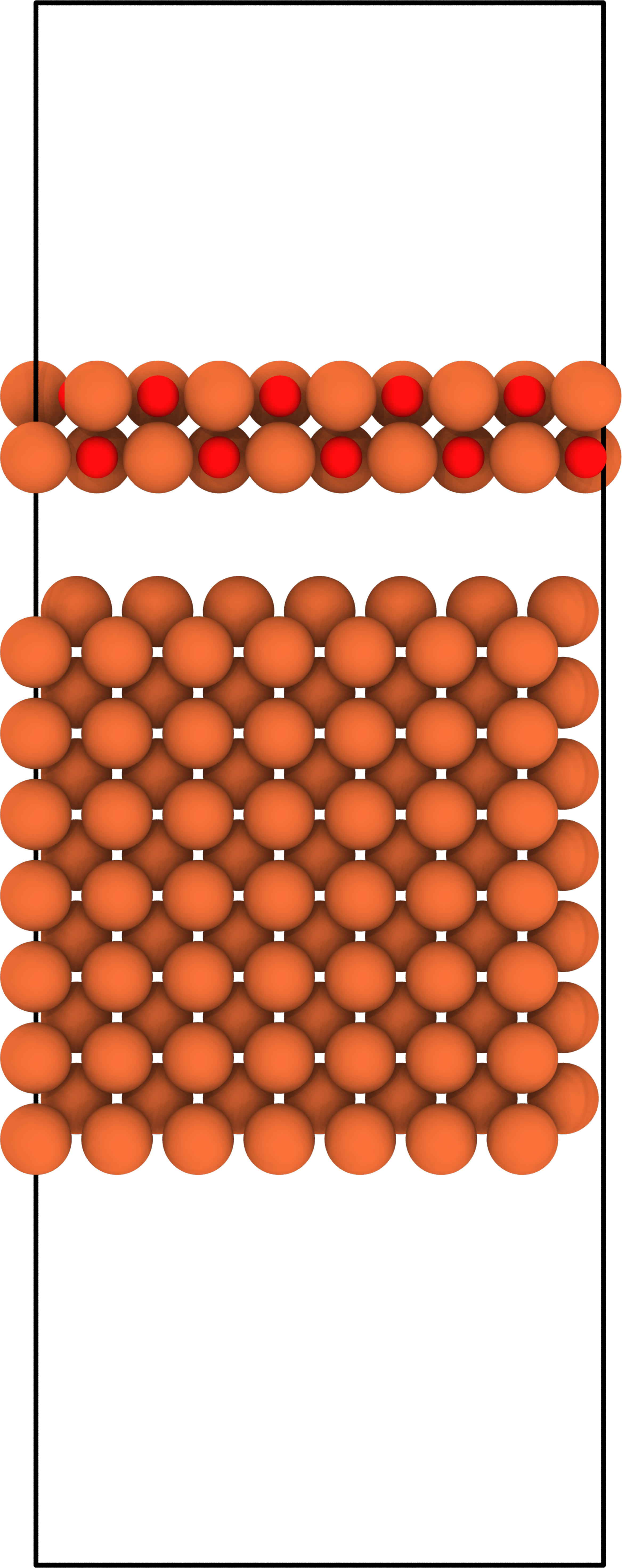}}
        \subfloat[$2$ ps]
         {\includegraphics[width=\figwf\columnwidth , height=10cm]{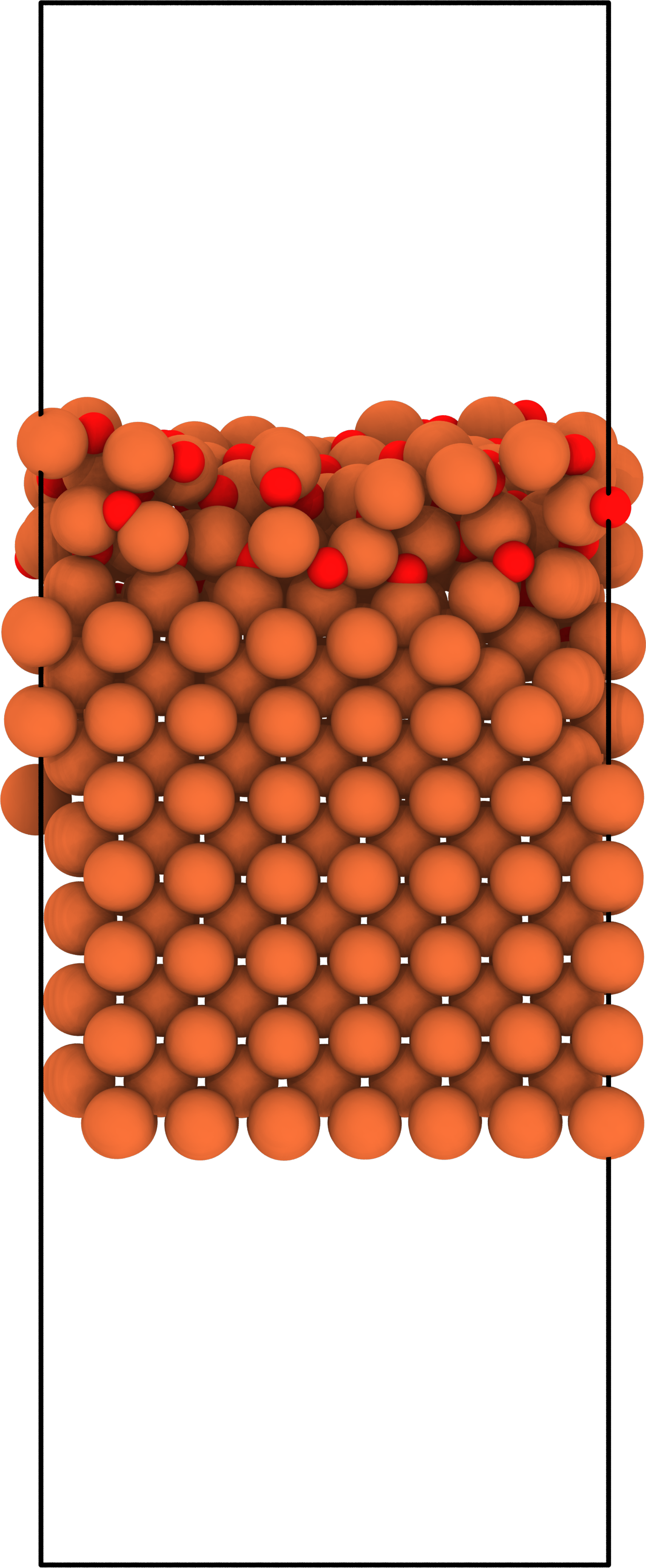}}
         \subfloat[$30$ ps]
         {\includegraphics[width=\figwf\columnwidth , height=10cm]{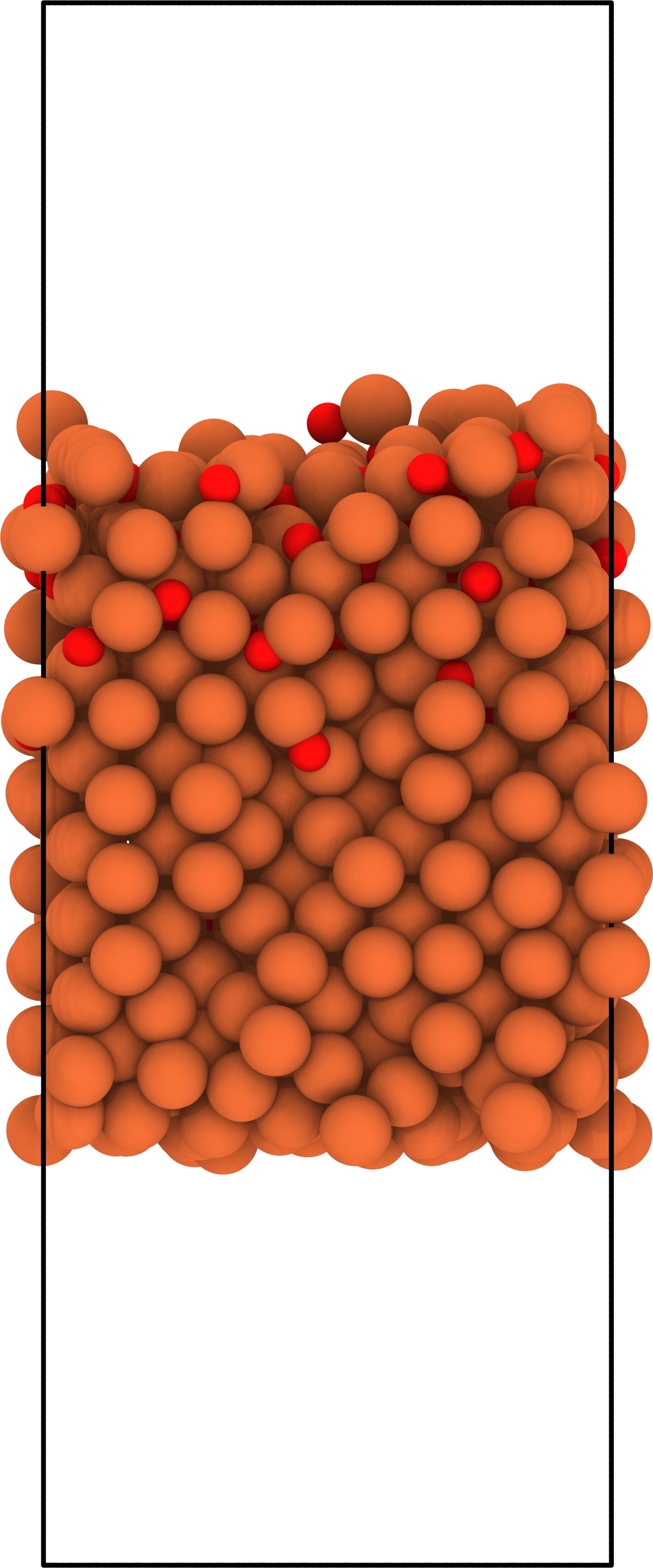}}
         \subfloat[$60$ ps]
         {\includegraphics[width=\figwf\columnwidth , height=10cm]{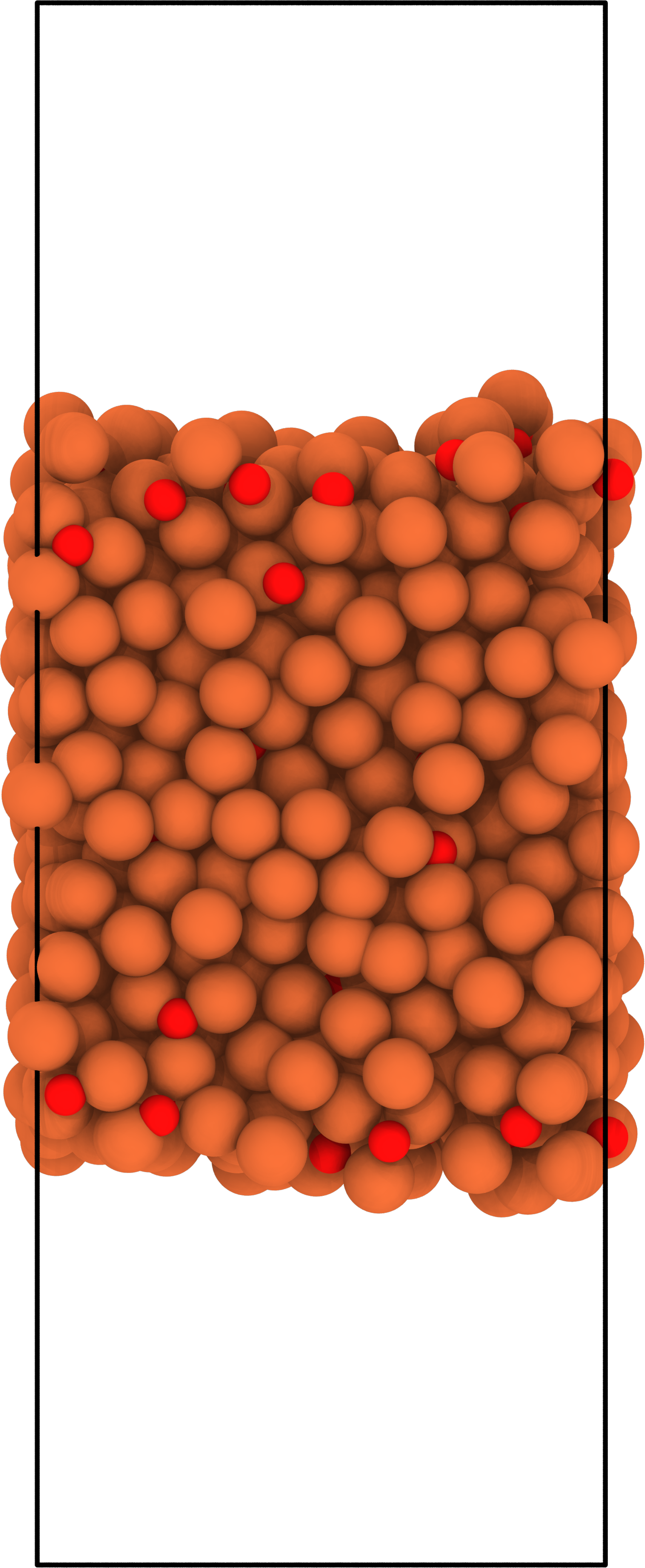}}
    \caption{Realization of an iron-oxide surface with $Z_\mathrm{O} = 0.11$ and $T_\mathrm{p} = 2000 \: \mathrm{K}$ for different moments in time during the preparation phase with heating strategy 1. The figures show a lateral surface, with the top surface used for the scattering gas molecule. Red particles represent oxygen atoms, while the brown particles are iron atoms. The sizes of the particles equal the atomic radii. \textcolor{black}{In (d), the normalized $z$-positions are indicated.} \textcolor{black}{CPK coloring is used to distinguish different chemical elements.}}%
     \label{fig:Surface_FeO_O2_Tp2000_ZO1}
\end{figure*}

\begin{figure}[h]
    \centering
    {\includegraphics[width=0.35\columnwidth]{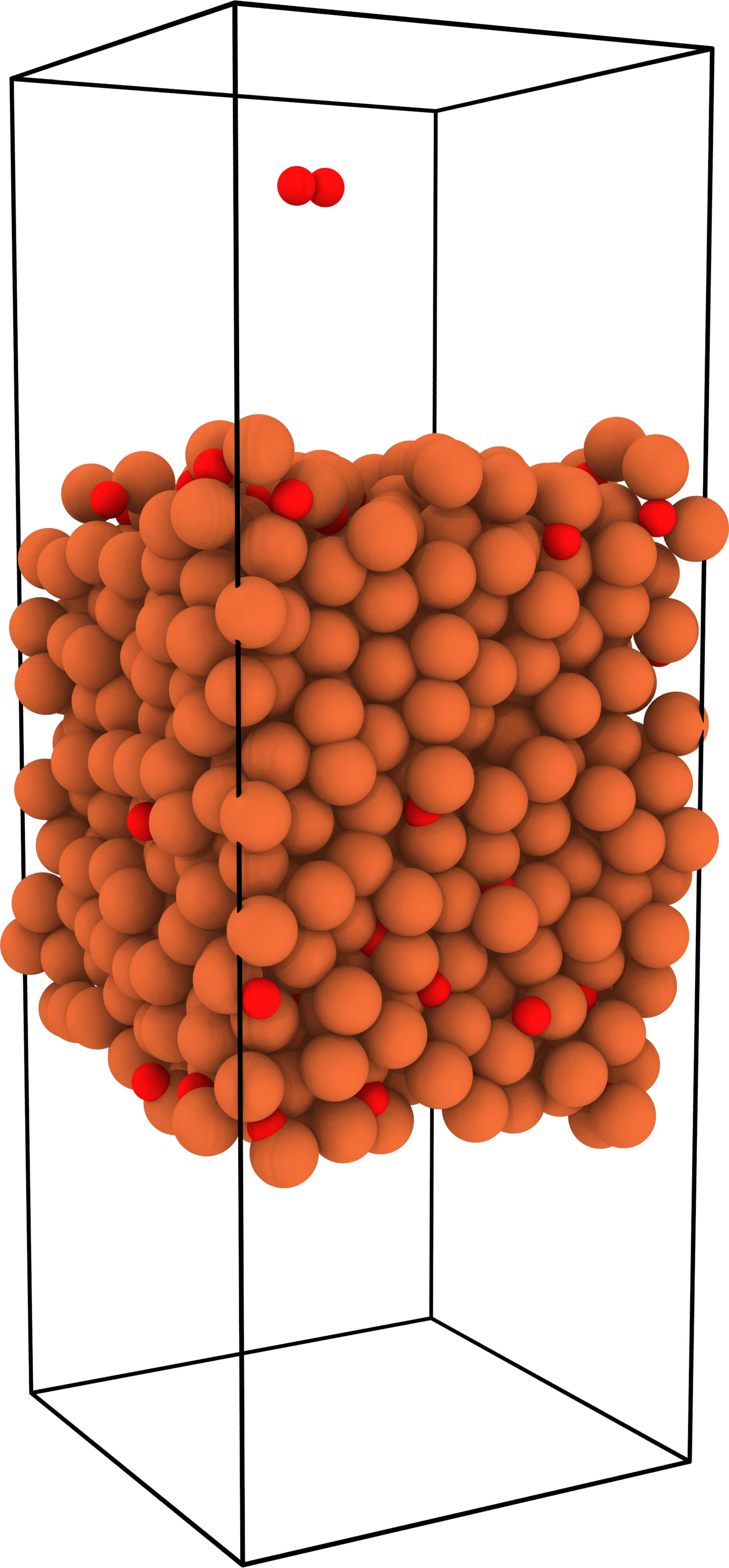}}
    \caption{Initial configuration showing an oxygen molecule (red particles) placed above a partially oxidized iron surface (brown particles) with $Z_\mathrm{O} = 0.11$ and $T_\mathrm{p} = 2000 \: \mathrm{K}$. The sizes of the particles equal the atomic radii. \textcolor{black}{CPK coloring is used to distinguish different chemical elements.} }%
     \label{fig:InitialConfigFeyOxO2}
\end{figure}

Before applying the Nose-Hoover thermostats, an $\mathrm{FeO}$ lattice is deposited in a specific ratio on top of a BCC lattice of $\mathrm{Fe}$ atoms. An initial distance of $4 \mathrm{\AA}$ between the two layers is introduced to ensure a natural mixing. The height of the $\mathrm{FeO}$ lattice compared to the $\mathrm{Fe}$ lattice is increased to increase the $Z_\mathrm{O}$ of the specific surface. Two different heating strategies are used to investigate the effect of the distribution of oxygen atoms over the surface:
\begin{enumerate}
    \item The same heating strategy previously discussed for the $\mathrm{Fe}$ surface is used, with a different heating strategy for a solid or liquid surface.
    \item An annealing process is employed for all the surfaces to enhance the mixing of the O and Fe atoms. The surfaces are warmed using the Nose-Hoover thermostat for 30 ps to $2800\:\mathrm{K}$, equilibrated for 30 ps at the same temperature and then gradually cooled down to the target temperature within 30 ps.
\end{enumerate}
 Figure \ref{fig:Surface_FeO_O2_Tp2000_ZO1} shows the preparation of an iron-oxide surface with $Z_\mathrm{O} = 0.11$ and $T_\mathrm{p} = 2000 \: \mathrm{K}$ via heating strategy 1. After the surface realization, an $\mathrm{O_2}$ molecule is located around $10\:\mathrm{\AA}$ above the surface, beyond the range of the potential well. Figure \ref{fig:InitialConfigFeyOxO2} shows the initial configuration used for the interaction between $\mathrm{Fe_xO_y}$ and $\mathrm{O_2}$.

Reactive molecular dynamics is used to simulate the interaction between $\mathrm{Fe_xO_y}$-$\mathrm{O_2}$. Reactive MD uses reactive force fields to accurately describe bond formation and breaking. ReaxFF \citep{Duin2001} is a bond order potential that describes the total energy of the system as
\begin{align}
    E_\mathrm{system} = E_\mathrm{bond} + E_\mathrm{over} + E_\mathrm{under} + E_\mathrm{val} + E_\mathrm{tor} \notag \\ + E_\mathrm{vdWaals} + E_\mathrm{Coulomb} + E_\mathrm{additional},
\end{align}
with $E_\mathrm{bond}$ the bond formation/breaking energy, $E_\mathrm{over}$ and $E_\mathrm{under}$ the over- and undercoordination energy penalties, $E_\mathrm{val}$ and $E_\mathrm{tor}$ are respectively the valence and torsion angle energies, $E_\mathrm{vdWaals}$ and $E_\mathrm{Coulomb}$ are the non-bonded van der Waals and Coulomb long-range interactions and $E_\mathrm{additional}$ are additional correction terms. The atomic charges are computed at every timestep using the charge equilibration method. A time step of $0.1 \: \mathrm{fs}$ is used, which is recommended for reactive MD simulations at high temperatures \citep{kritikos2022}. The timestep is held sufficiently small to capture all reaction events at high temperatures. \textcolor{black}{Six different surfaces, each with different initial velocities, are generated to obtain a statistically meaningful set of data.}  \textcolor{black}{Table \ref{Tab:interactions} lists the inter-atomic potentials that are used in this work to model the various inter-atomic interactions.}

 \begin{table*}[ht]
\centering
    \caption{\textcolor{black}{Inter-atomic potentials used for the different atom-atom pair for either the $\mathrm{Fe}$ - $\mathrm{N_2}$ or $\mathrm{Fe_xO_y}$ - $\mathrm{O_2}$ interaction.}}
\resizebox{\textwidth}{!}{\begin{tabular}{cccc}
 \hline
Case & Atom-atom pair & Inter-atomic potential & Reference\\
 \hline
$\mathrm{Fe}$ - $\mathrm{N_2}$ & $\mathrm{Fe}$ - $\mathrm{Fe}$ & EAM potential   & \cite{Zhou2004}\\
$\mathrm{Fe}$ - $\mathrm{N_2}$ & $\mathrm{N}$ - $\mathrm{N}$ & Rigid rotor & \cite{Decius1996} \\
$\mathrm{Fe}$ - $\mathrm{N_2}$ & $\mathrm{Fe}$ - $\mathrm{N_2}$ & Morse potential  & \cite{Sipkens2014}\\
$\mathrm{Fe_xO_y}$ - $\mathrm{O_2}$ & $\mathrm{Fe}$ - $\mathrm{Fe}$ & ReaxFF  & \cite{ReaxFF2010}\\
$\mathrm{Fe_xO_y}$ - $\mathrm{O_2}$ & $\mathrm{O}$ - $\mathrm{O}$ & ReaxFF    & \cite{ReaxFF2010}\\
$\mathrm{Fe_xO_y}$ - $\mathrm{O_2}$ & $\mathrm{Fe}$ - $\mathrm{O}$ & ReaxFF    & \cite{ReaxFF2010}\\
 \hline
\end{tabular}}
    \label{Tab:interactions}
\end{table*}
\section{Results of the molecular dynamics simulations}\label{sec:results}
The results for the mass and thermal accommodation coefficients determined from MD simulations are discussed below. \textcolor{black}{It is important to note here that the MD simulation results are to some extent limited by the available (reactive) force fields. In addition, results from different force fields could differ quantitatively. Therefore, care must be taken in the selection of the force field.}
\textcolor{black}{The effect of the available reactive force fields on the predicted thermal expansion is examined and presented in the Supplementary Material. As a result, it is concluded that ReaxFF proposed by \cite{ReaxFF2010} is suited to reproduce an $\mathrm{FeO}$ and $\mathrm{Fe_3O_4}$ surface. Therefore, this reactive force field is used in this work.} 
\subsection{Mass accommodation coefficients}\label{sec:macresults}
The mass accommodation coefficient for iron with oxygen is investigated for different initial oxidation stages, ranging from $Z_\mathrm{O} = 0$ to $Z_\mathrm{O} = 0.57$, and three different surface temperatures, namely, $T_\mathrm{p} = 1500$, $2000$ and $2500~\mathrm{K}$, of which the latter two are in the liquid-phase regime. \textcolor{black}{A gas temperature of $T_\mathrm{g} = 300\: \mathrm{K}$ is used.} \textcolor{black}{The distance between the two $\mathrm{O}$ atoms of the incident oxygen molecule is used as a metric for the MAC. If the inter-atomic distance becomes larger than $1.2 \times$ the initial bond length, the incident oxygen molecule is considered as dissociated, and chemically absorbed into the iron(-oxide) surface.}

\begin{figure*}
    \centering
        \subfloat[$Z_\mathrm{O} = 0.22$]
         {\includegraphics[width=\figwdouble\columnwidth, clip]{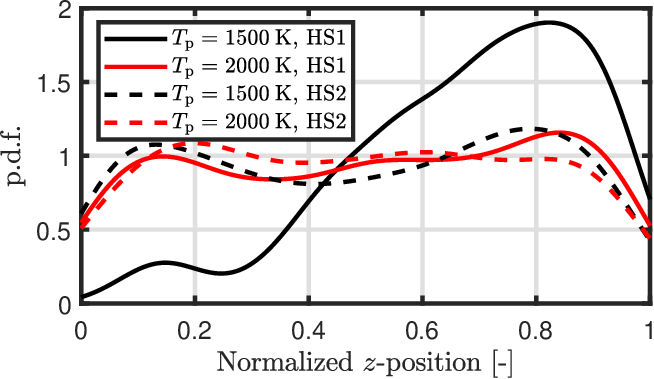}}
        \subfloat[$Z_\mathrm{O} = 0.45$]
         {\includegraphics[width=\figwdouble\columnwidth, clip]{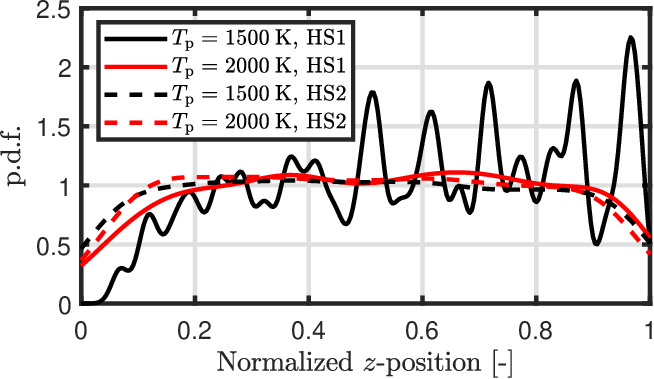}}
    \caption{Probability density function of oxygen atoms $z$-position for two different heating strategies, $T_\mathrm{p}$ and $Z_\mathrm{O}$. }%
     \label{fig:pdfo_O}
\end{figure*}

As discussed in Section \ref{sec:FexOy_O2_interaction}, two different heating strategies are employed to investigate the effect of the non-uniform distribution of oxygen atoms in the surface. Figure \ref{fig:pdfo_O} shows the probability density of oxygen atoms as a function of normalized $z$-position \textcolor{black}{(as shown in Fig. \ref{fig:Surface_FeO_O2_Tp2000_ZO1}d)}. For $T_\mathrm{p} = 2000 \: \mathrm{K}$, the oxygen atoms are characterized by a relatively uniform distribution, independent of the heating strategy. This lack of dependency is, however, not the case for $T_\mathrm{p} = 1500 \: \mathrm{K}$. With heating strategy~1 (HS1), the oxygen atoms are not uniformly distributed---a higher concentration is observed near the top surface. Because of the heating rate of HS1, not all of the oxygen atoms are sufficiently diffused to reach a uniform distribution. As a result, more oxygen atoms are near the surface. When the annealing process of HS2 is used, a spatially more uniform distribution of oxygen atoms is obtained.

\begin{figure*}
    \centering
        \subfloat[]
         {\includegraphics[width=\figwdouble\columnwidth, clip]{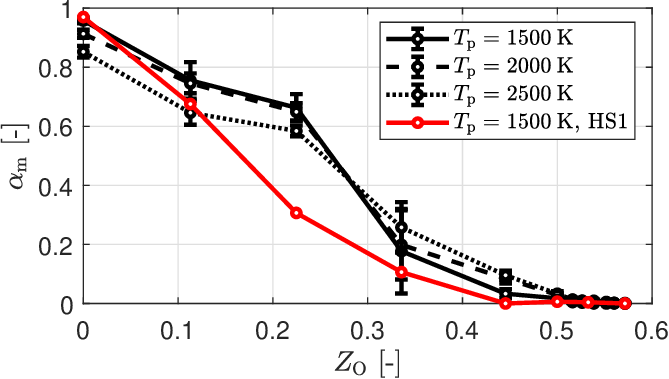}}
        \subfloat[]
         {\includegraphics[width=\figwdouble\columnwidth, clip]{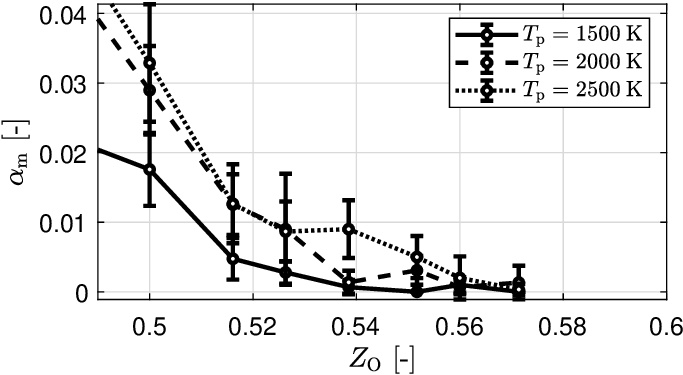}}
    \caption{The mass accommodation coefficient of oxygen as a function of initial oxidation stage at three different surface temperatures and \textcolor{black}{$T_\mathrm{g} = 300\: \mathrm{K}$}. A detailed view around $Z_\mathrm{O} = 0.5$ is shown in (b).}%
     \label{fig:MAC_Tg300K}
\end{figure*}

Figure \ref{fig:MAC_Tg300K}a shows the MAC of oxygen as a function of initial oxidation stage for the three different surface temperatures, obtained with HS2, while Fig. \ref{fig:MAC_Tg300K}b shows a detailed view around $Z_\mathrm{O} = 0.57$. To show the effect of a higher oxygen concentration at the top surface, the MAC obtained with HS1 for $T_\mathrm{p} = 1500 \: \mathrm{K}$ is added to Figure \ref{fig:MAC_Tg300K}a as a red line. With a homogeneously distributed surface obtained via HS2, the MAC barely depends on the surface temperature and decreases as a function of $Z_\mathrm{O}$. \textcolor{black}{A fast decrease in MAC with increasing $Z_\mathrm{O}$ is observed if $Z_\mathrm{O} < 0.5$, whereas the decrease of MAC becomes weaker for $Z_\mathrm{O} > 0.5$. This result indicates that once the particle reaches stoichiometric $\mathrm{FeO}$, it becomes more difficult to absorb oxygen}.

As depicted in Figure \ref{fig:MAC_Tg300K}a with the red line, the MAC decreases significantly if there are more oxygen atoms near the surface. This accumulation at the top surface prevents new oxygen atoms from being accommodated, indicating that the mass accommodation coefficient is affected by the local oxygen concentration near the surface. \textcolor{black}{Thus, in this region the MAC is affected by a separation of time scales: The time period over which a gas molecule interacts with the surface is much shorter than the time period over which local oxygen concentration near the surface changes}. Therefore, if the iron particle does not have a spatially uniform composition, internal transport may limit the oxidation rate of iron particles.

\subsection{Thermal accommodation coefficients}
\subsubsection{$\mathrm{Fe}$ - $\mathrm{N_2}$ interactions}


The TAC for the $\mathrm{Fe}$ - $\mathrm{N_2}$ interaction is investigated for different surface temperatures and \textcolor{black}{a gas temperature of $T_\mathrm{g} = 300\: \mathrm{K}$}. An initially rough surface was created by projecting iron atoms on the initial smooth lattice. \textcolor{black}{It is found that the TAC for the $\mathrm{Fe}$ - $\mathrm{N_2}$ interaction is almost independent of temperature and equals $\alpha_\mathrm{T} = 0.17$.} These values are consistent with the values experimentally obtained by \cite{Sipkens2014}, who reported a value of $0.10$ and $0.17$. \textcolor{black}{The effect of surface roughness on the TAC is discussed in the Supplementary Material.}

In addition to the surface temperature of iron, the temperature of the surrounding gas can change significantly during iron combustion. \cite{Daun2009} and \cite{Mane2018} investigated the effect of gas temperature on the TAC of nitrogen with soot and hydrogen with aluminum, respectively. They showed that the TAC is almost independent of gas temperature and primarily influenced by surface roughness and gas molecular weight. Based on the current MD results, a recommended value for the TAC of the interaction between $\mathrm{Fe}$ with $\mathrm{N_2}$ equals $0.17$. 

\subsubsection{$\mathrm{Fe_xO_y}$ - $\mathrm{O_2}$ interactions}
\begin{figure}[h]
    \centering
    {\includegraphics[width=\figwsingle\columnwidth, clip]{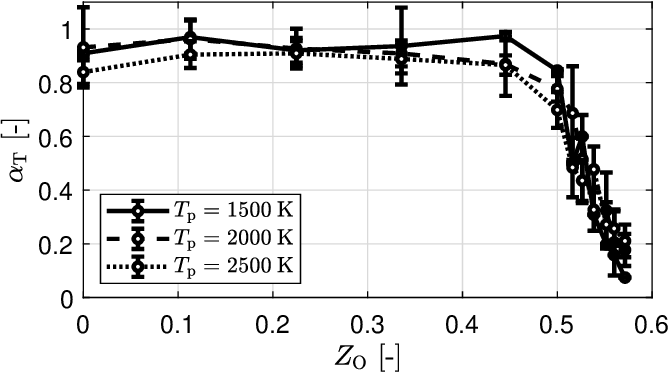}}
    \caption{Total thermal accommodation coefficients of the $\mathrm{Fe_xO_y}$-$\mathrm{O_2}$ interaction as a function of $Z_\mathrm{O}$ at three different surface temperatures \textcolor{black}{, with $T_\mathrm{g} = 300\: \mathrm{K}$.}}%
     \label{fig:TAC_O2_Tg300K}
\end{figure}

The oxygen molecules that do not stick to the surface during $\mathrm{Fe_xO_y}$-$\mathrm{O_2}$ interactions still contribute to the TAC. Figure \ref{fig:TAC_O2_Tg300K} shows the total thermal accommodation coefficients of the $\mathrm{Fe_xO_y}$-$\mathrm{O_2}$ interaction as a function of $Z_\mathrm{O}$ at three different surface temperatures and \textcolor{black}{a gas temperature of $T_\mathrm{g} = 300\: \mathrm{K}$}. When the oxidation degree of the surface is low, the TAC remains close to unity but decreases sharply to $0.2$ once $Z_\mathrm{O} > 0.5$. \textcolor{black}{This trend can be explained by means of the number of hits that the incoming oxygen molecule has with the iron(-oxide) surface. With an increasing $Z_\mathrm{O}$, the probability in number of hits decreases since the incoming oxygen is repelled by the surface. In other words, in a case with a larger $Z_\mathrm{O}$, the oxygen molecule spends less time in contact with the surface, and therefore does not have the opportunity to equilibrate with the surface, resulting in a lower thermal accommodation coefficient.}
Note that, if $Z_\mathrm{O}$ is small, and thus MAC is large, the amount of scattered oxygen atoms over which the TAC is calculated is low. Since the total thermal accommodation coefficient is calculated with Equation~\eqref{eq:TACTotal}, however, this uncertainty in the low $Z_\mathrm{O}$ regime could be neglected due to the small contribution of $\alpha_\mathrm{T,O_2}$ given by the large MAC number. 

\subsection{Discussion of the MD results}\label{sec:discussion_md}
A limitation of \textcolor{black}{the herein used molecular dynamics} approach to determine the TAC and MAC is the assumption of a clean surface. Within experiments, surfaces could be (partly) covered by a layer of gas molecules. \cite{Song1987} proposed a semi-empirical correlation for the TAC of engineering surfaces, which takes into account this absorded layer. They state that the correlation is general and can be used for any combination of gases with a solid surface over a wide temperature range. However, \cite{Song1987} state that a common strategy to create an adsorption-free surface within experiments is to heat the surface to high temperatures above $1000 \: \mathrm{K}$ to desorb all of the impurities in the surface. The present work is mainly focused on the accommodation coefficients in the liquid-phase regime. Since the current MD simulations better represent the condition of a liquid-phase surface rather than a solid-phase surface with intrinsic roughness, the resulting accommodation coefficients in the temperature range above the melting points are likely more accurate. \textcolor{black}{\cite{Philyppe2022} used the correlation for the TAC as proposed by \cite{Song1987} to investigate the ignition of iron particles in the Knudsen transition regime. A comparison between the TAC of the $\mathrm{Fe}$-$\mathrm{N_2}$ interaction obtained with the MD simulations compared to the semi-empirical correlation of \cite{Song1987} is presented in the Supplementary Material. The TAC of \cite{Song1987} is about three times larger than the TAC obtained with the molecular dynamics simulations.} 

\cite{Nejad20202} performed molecular dynamics simulations to investigate the influence of gas-wall and gas-gas interaction on different accommodation coefficients. They used a parallel wall approach to determine the accommodation coefficients, indicating that an intermediate Knudsen number is modeled. They showed that \textcolor{black}{neglecting the gas-gas interaction results in lower TAC values with respect to the parallel wall approach: The TAC could be around 1.5 times larger if the gas-gas interaction effect were included \citep{Nejad20202}}. However, in the two-layer model used in this work, a free-molecular regime is assumed within the Knudsen layer, implying that each incoming gas molecule does not interact with each other. Therefore, the accommodation coefficients obtained \textcolor{black}{when neglecting the gas-gas interactions} are consistent with the used two-layer model. 


\subsection{Implementation of the MD results in the single iron particle combustion model}
The TAC and MAC obtained from the MD simulations are used in the single iron particle combustion model. In the iron particle model, it is assumed that the particle has a homogeneous mixture, such that there is no internal gradient in oxygen concentration. Therefore, the TAC and MAC values for $\mathrm{Fe_xO_y}$-$\mathrm{O_2}$ interactions are used from HS2, which are a function of $Z_\mathrm{O}$ and nearly independent of the surface temperature. In Table \ref{tab:TACMACFeO2} the TAC and MAC for the $\mathrm{Fe_xO_y}$ - $\mathrm{O_2}$ interactions are listed and averaged over the three surface temperatures.
Within the iron particle combustion model, these values are used for linear interpolation. However, if no further oxidation is modeled after reaching the stoichiometry of $\mathrm{FeO}$, a linear fit (\textcolor{black}{$\mathrm{MAC_{Z_\mathrm{O} < 0.5}} = -1.8Z_\mathrm{O} + 0.9$}) is used for the MAC such that $\alpha_\mathrm{m} = 0$ if $Z_\mathrm{O} = 0.5$. . Equation \eqref{eq:TACTotal} describes the total TAC with $\alpha_\mathrm{T,N_2} = 0.17$ and $\alpha_\mathrm{T,O_2} = f(Z_\mathrm{O})$. 

\begin{table*}
\centering
    \caption{TAC and MAC for the $\mathrm{Fe_xO_y}$ - $\mathrm{O_2}$ interactions, averaged over the three phase temperatures.}
\resizebox{\textwidth}{!}{\begin{tabular}{ccccccccccccc}
\hline
$Z_\mathrm{O}$ & $0$ & $0.11$ & $0.22$ & $0.34$ & $0.45$ & $0.5$ & $0.52$ & $0.53$ & $0.54$ & $0.55$  & $0.56$  & $0.57$
\\
\hline
$\alpha_\mathrm{m}$ & $0.908$ & $0.716$ & $0.632$ & $0.211$ & $0.071$ & $0.027$ & $0.010$ & $0.007$ & $0.004$ & $0.003$ & $0.001$ & $0.001$\\
$\alpha_\mathrm{T}$ & $0.893$ & $0.946$ & $0.919$ & $0.911$ & $0.903$ & $0.774$ & $0.557$ & $0.517$ & $0.371$ & $0.265$ & $0.207$ & $0.154$\\
\hline
\end{tabular}}

    \label{tab:TACMACFeO2}
\end{table*}

\section{Results of single iron particle combustion simulations}
In this section, the results of the MD-informed Knudsen model for a single iron particle burning in an $\mathrm{O_2}$-$\mathrm{N_2}$ atmosphere are presented. An initial particle temperature just above the ignition temperature, $T_{\mathrm{p,0}} = 1100\:  \mathrm{K}$ \citep{Mi2022}, is considered.

\subsection{Combustion behavior}
First, the effect of the new model by only considering the first oxidation stage up to $Z_\mathrm{O} = 0.5$ is investigated. The initial conditions are chosen such that the laser-ignited experiments performed by \cite{Ning2020} are mimicked. A cold gas of $T_{\mathrm{g,0}} = 300 \:  \mathrm{K}$ at $1 \: \mathrm{atm}$ is considered. The temperature profiles are shifted such that the particle temperature equals $T_{\mathrm{p,0}} = 1500\: \mathrm{K}$ at $t = 0 \: \mathrm{ms}$ \citep{Ning2021}. 

\begin{figure*}
    \centering
        \subfloat[]
         {\includegraphics[width=\figwdouble\columnwidth, clip]{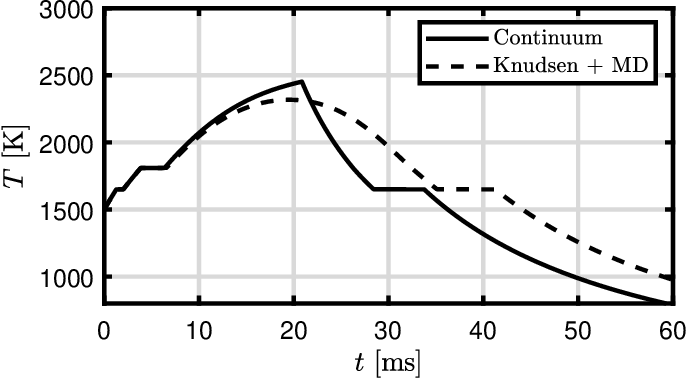}}
        \subfloat[]
         {\includegraphics[width=\figwdouble\columnwidth, clip]{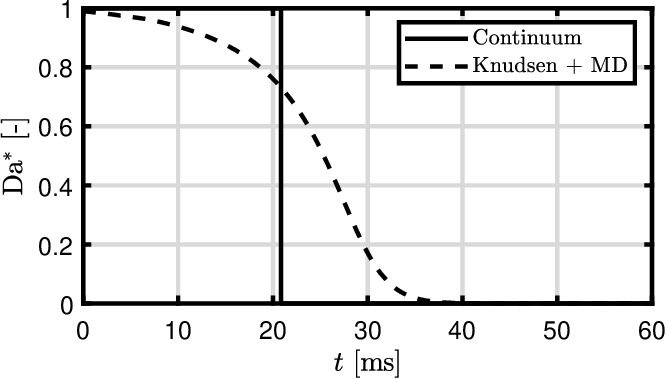}}
\caption{(a) Temperature profile and (b) normalized Damköhler number for an iron particle of $50\:$\textmu m burning at 21\% oxygen concentration. The continuum model is compared to the MD-informed Knudsen model.} \label{fig:Tvst_MD_Cont_dp50_XO21}
\end{figure*}

\begin{figure*}
    \centering
        \subfloat[]
         {\includegraphics[width=\figwdouble\columnwidth, clip]{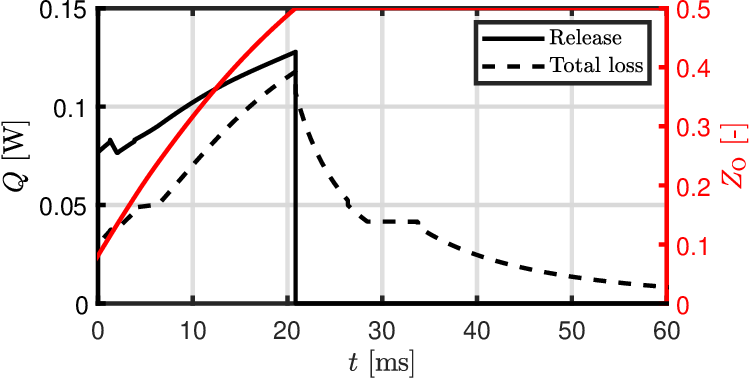}}
        \subfloat[]
         {\includegraphics[width=\figwdouble\columnwidth, clip]{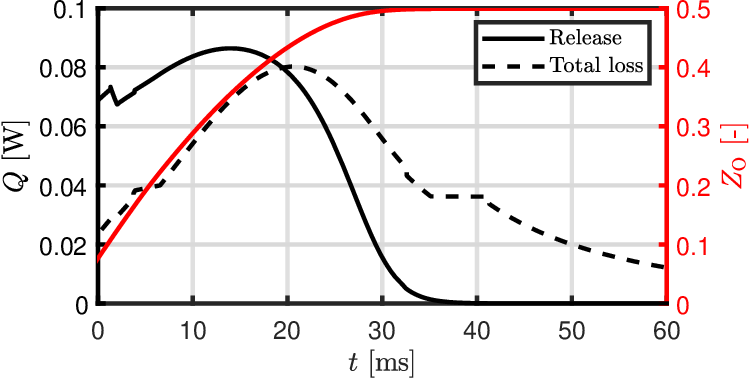}}\\
\caption{Heat release and heat loss rates (left axis) and $Z_\mathrm{O}$ (right axis) for an iron particle of $50\:$\textmu m burning at 21\% oxygen concentration. The results are shown for (a) the continuum model and (b) the MD-informed Knudsen model.} \label{fig:YQKnudsenCont}
\end{figure*}

Figure \ref{fig:Tvst_MD_Cont_dp50_XO21}a shows the comparison of the temperature profiles between the continuum model and the MD-based Knudsen model for a $50$\textmu m particle burning in $X_\mathrm{O_2} = 0.21$. The temperature \textit{vs}. time curve for the MD-informed Knudsen model changes significantly compared to the previously used continuum model. This difference can be explained by examining the $Z_\mathrm{O}$ value and heat transfer rates plotted in Figure \ref{fig:YQKnudsenCont}. In the continuum model, the maximum temperature was located at the position where $Z_\mathrm{O} = 0.5$ is reached. At that time, the available iron is completely oxidized, and therefore the heat release rate immediately drops to zero. With the MD-based Knudsen model, this behavior changes and $Z_\mathrm{O} = 0.5$ is reached after the peak temperature. Since the rate of oxidation slows down, as the MAC decreases with an increasing oxidation stage, the rate of heat loss exceeds the rate of heat release upon reaching the maximum temperature of the particle, \textcolor{black}{while the particle is not yet oxidized to $Z_\mathrm{O} = 0.5$}.

With the continuum model, it was discussed that the particle burns in a regime limited by the external diffusion of oxygen prior to the maximum temperature. One can derive a normalized Damköhler number $\mathrm{Da^*}$. If $\mathrm{Da^*}$ is close to zero, the particle burns in a kinetic- (or chemical-) absorption-limited regime, and if it is close to unity, the particle burns in an external-diffusion-limited regime. Figure \ref{fig:Tvst_MD_Cont_dp50_XO21}b shows the normalized Damköhler number for the same configuration. The normalized Damköhler number of the continuum model is determined according to the definition of \cite{Hazenberg2020}. For the Knudsen + MD model, the normalized Damköhler number is defined as
\begin{equation}
    \mathrm{Da^*} = 1 - \frac{\alpha_\mathrm{m} X_\mathrm{O_2,\delta}}{X_\mathrm{O_2,g}},
\end{equation}
where $\alpha_\mathrm{m} X_\mathrm{O_2,\delta}$ denotes the molar fraction of oxygen at the particle surface $X_\mathrm{O_2,p}$. As discussed before, with the continuum model the particle burns completely in an external-diffusion-limited regime prior to the maximum temperature. This behavior changes in the MD-informed Knudsen model: Due to the decreasing MAC value with an increasing oxidation stage, the particle burns in an intermediate regime. 


\subsection{Comparison with experimental results}
In \cite{Thijs2022,ThijsPCI_2022}, only the first stage of combustion, which is the oxidation to $Z_\mathrm{O} = 0.5$, was investigated. Due to this assumption, an inert cooling stage was observed after the peak temperature. However, \cite{Choisez2022} investigated combusted iron powders and discovered that it primarily consisted of a magnetite and hematite mixture, indicating a $Z_\mathrm{O}$ greater than $0.5$. With the MD-informed Knudsen model, the oxidation beyond $Z_\mathrm{O} = 0.5$ can be included in the combustion of a single particle. 

The results of the MD-informed Knudsen model are compared with two sets of experimental data. First, the model is compared to the laser-ignited experiments of \cite{Ning2021} wherein the particles burn in air at $300 \: \mathrm{K}$. Then, the new temperature curve is compared to the drop-tube experiments of \cite{Panahi2022} wherein the particles burn in varying oxygen concentrations at $1350 \: \mathrm{K}$. The experimental data are averaged over multiple independent single-particle measurements to obtain a smooth curve.

\subsubsection{Comparison with Ning et al.}
\begin{figure*}
    \centering
        \subfloat[]
         {\includegraphics[width=\figwdouble\columnwidth, clip]{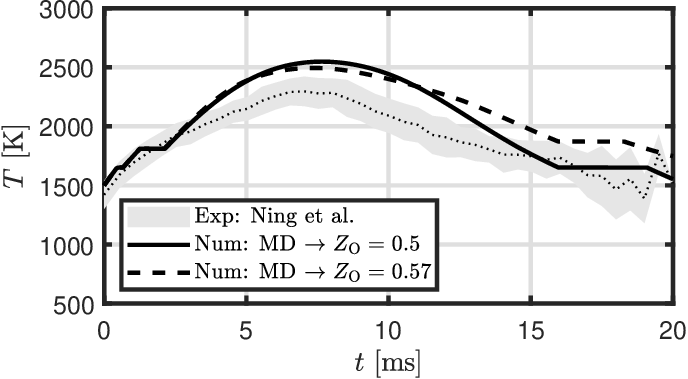}}
        \subfloat[]
         {\includegraphics[width=\figwdouble\columnwidth, clip]{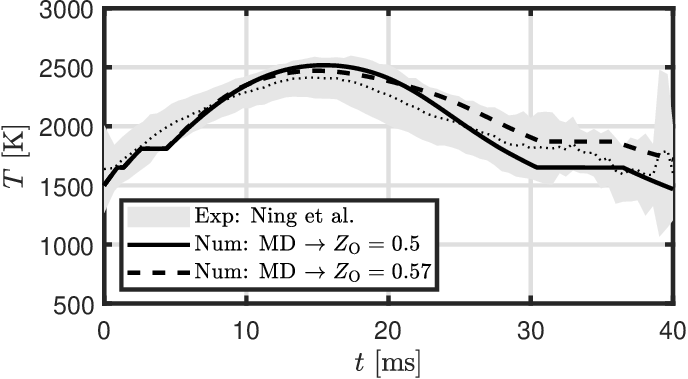}}
 \caption{Temperature profile for an iron particle of (a) $34$\textmu m and (b) $50$\textmu m burning at 26\% oxygen concentration, with and without further oxidation beyond $Z_\mathrm{O} = 0.5$. The dotted line and gray area are the mean and standard deviation of measurements obtained with the setup of \cite{Ning2020}, respectively. }\label{fig:Knudsen_furtherOx_Tp_XO26_Exp}
\end{figure*}

Figure \ref{fig:Knudsen_furtherOx_Tp_XO26_Exp} shows the temperature profiles for the MD-informed Knudsen model with and without further oxidation beyond $Z_\mathrm{O} = 0.5$ for a $34$\textmu m and $50$\textmu m particle burning in air with $X_\mathrm{O_2} = 0.26$. The dotted line and gray area in Figure \ref{fig:Knudsen_furtherOx_Tp_XO26_Exp} are the mean and standard deviation of the experimentally obtained temperature profiles, respectively.

The particle temperature for smaller particles is overestimated. Overall, the temperature curve obtained with the MD-informed Knudsen model shows a better agreement with the experimentally obtained temperature curve in contrast to the \textcolor{black}{less smooth curve obtained with the external-diffusion limited continuum-model prediction, as shown in Figure \ref{fig:Tvst_MD_Cont_dp50_XO21}}. The sharp transition at the peak temperature is now a smooth curve. The inclusion of the oxidation beyond $Z_\mathrm{O} = 0.5$ results in a higher temperature in the tail of the curve. Instead of inert cooling after $Z_\mathrm{O} = 0.5$, a reactive cooling regime is observed. The new numerically obtained slopes after the peak temperature qualitatively better agree with the experimental measurement during the cooling stage. 

\subsubsection{Comparison with Panahi et al.}
\begin{figure*}
    \centering
        \subfloat[]
         {\includegraphics[width=\figwdouble\columnwidth, clip]{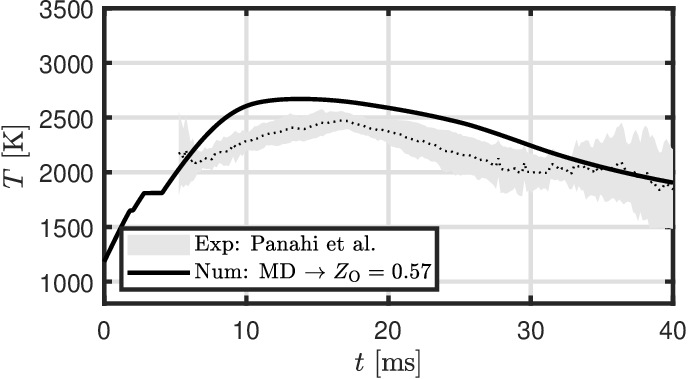}}
        \subfloat[]
         {\includegraphics[width=\figwdouble\columnwidth, clip]{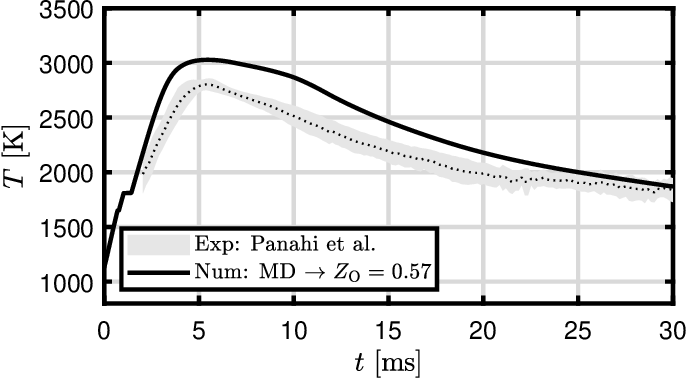}}\\
         \subfloat[]
         {\includegraphics[width=\figwdouble\columnwidth, clip]{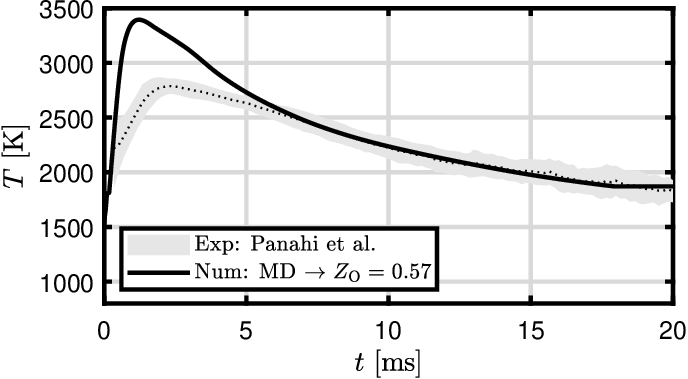}}
 \caption{Temperature profile for an iron particle of $49\:$\textmu m burning at (a) 21\%, (b) 50\% and (c) 99 \% oxygen concentration. The dotted line and gray area are the mean and standard deviation of measurements obtained with the setup of \cite{Panahi2022}, respectively. } \label{fig:Panahi_et_al_dp49}
\end{figure*}

Figure \ref{fig:Panahi_et_al_dp49} shows the temperature profiles for the MD-informed Knudsen model with further oxidation beyond $Z_\mathrm{O} = 0.5$ for a $49$\textmu m particle burning in air with $X_\mathrm{O_2} = 0.21$, $0.5$ and $0.99$. The dotted line and gray area are the mean and standard deviation of experimentally obtained temperature profiles, respectively, which are calculated via the setup described by \cite{Panahi2022}. Note that the effect of the Stefan flow correction on the evaporation rate is not taken into account. The data of \cite{Panahi2022} are time-shifted so that the first experimental data point approximates the numerical temperature.

Although the model overestimates the particle temperature at the two higher oxygen concentrations, the agreement after the maximum temperature is reasonable. The reactive cooling slope which is observed after the maximum particle temperature seems to match the experimentally observed slope. This qualitative agreement implies that the oxidation process of the particle continues after the maximum particle temperature is reached. 

A possible explanation for the overestimation of the particle temperature at higher oxygen concentrations could be due to \textcolor{black}{assumption of an} infinitely fast internal transport for cases with these high oxygen concentrations. As shown in Section \ref{sec:macresults}, the mass accommodation coefficient significantly decreases when the particle does not have a homogeneous composition, but a higher concentration of oxygen atoms near the surface. Since for high oxygen concentrations in the gas, the external diffusion of oxygen is fast, the diffusion of oxygen in the condensed phase could be rate-limiting. With an increasing oxidation stage $Z_\mathrm{O}$ and an increased oxygen concentration $X_\mathrm{O_2}$ in the gas phase, internal transport could limit the absorption rate of new oxygen molecule, and therefore limit the maximum particle temperature. \textcolor{black}{Another possible explanation is that the current point-particle model is not valid in case of these extreme conditions. While it is already taken into account that the evaporated gaseous iron(-oxide) will further oxidize, and therefore inhibit the oxygen diffusion towards the particle, more components must be taken into account. The dense nano-particle cloud surrounding the particle can to some extent affect the boundary layer. Therefore, the validity of Equation \ref{eq:mdot_cont} must be reconsidered. On the one hand, including gas-phase reactions will enhance the evaporating flux, since gaseous iron will immediately oxidize and is therefore consumed. On the other hand, a strong Stefan flow induced by the surface chemisorption of oxygen would inhibit evaporation. Furthermore, how the nano-particle cloud affects the heat transfer in the boundary layer is also not taken into account in the current point-particle model. In \cite{Thijs2022} boundary-layer resolved simulations are performed to validate the current point-particle model, but only simulations for relatively low oxygen concentrations are performed, without gas-phase kinetics. To validate the current point-particle model, boundary-layer resolved simulations at these high oxygen concentrations must be performed to investigate the validity of the current point-particle model under these extreme conditions.}

\subsection{Effect of $\alpha_\mathrm{T}$ and $\alpha_\mathrm{m}$}
The results of the molecular dynamics simulations are dependent on the accuracy of the inter-atomic potentials and, perhaps, also on the configuration. Related changes in the TAC and MAC have an effect on the temperature profile during combustion. In this section, the TAC and MAC are varied independently to investigate the effect on the combustion behavior.

\begin{figure*}
    \centering
        \subfloat[]
         {\includegraphics[width=\figwdouble\columnwidth, clip]{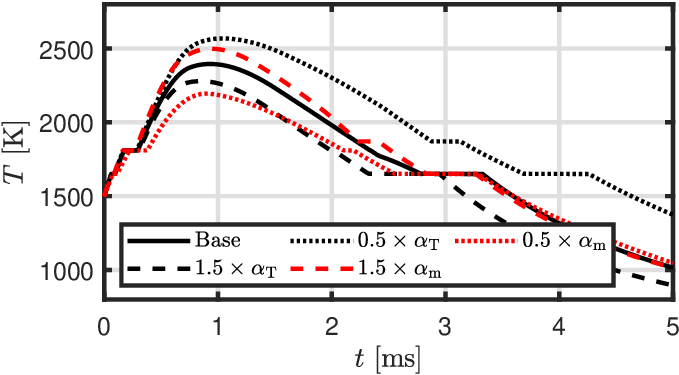}}
        \subfloat[]
         {\includegraphics[width=\figwdouble\columnwidth, clip]{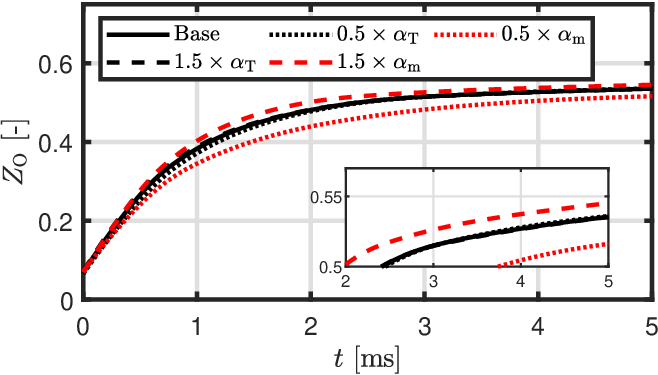}}\\     
           \subfloat[]
           {\includegraphics[width=\figwdouble\columnwidth, clip]{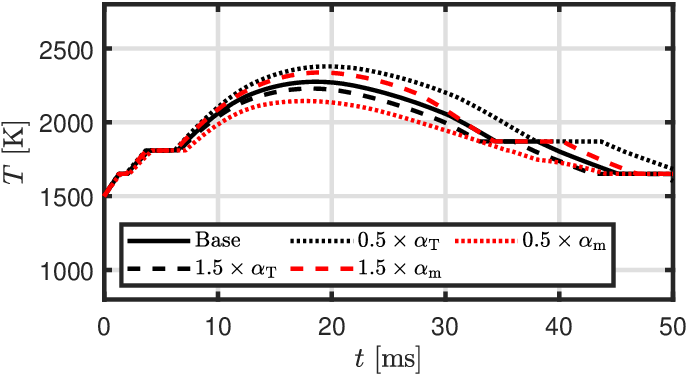}}
        \subfloat[]
         {\includegraphics[width=\figwdouble\columnwidth, clip]{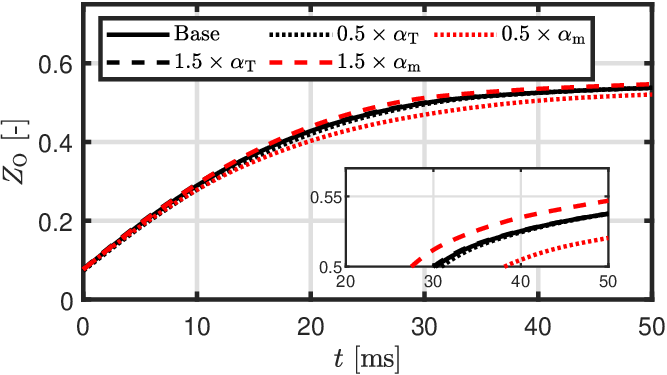}}
\caption{The effect of changes in $\alpha_\mathrm{T}$ and $\alpha_\mathrm{m}$ for a iron particle of \textcolor{black}{(a) + (b) $10\:$\textmu m and } (b) + (c) $50\:$\textmu m burning at 21\% oxygen concentration. (a) + (c) The particle temperature and (b) + (d) $Z_\mathrm{O}$. } \label{fig:EffectTACMAC_dp50um_xO26_Tp}
\end{figure*}

\textcolor{black}{As discussed in Section \ref{sec:discussion_md}, the TAC could be around 1.5 times larger if gas-gas interactions are taken into account. Although neglecting the gas-gas interaction in the current case is justified, this extreme variation of 50\% is used in this section to independently vary the TAC and MAC, with a maximum value of $1$}. Figure~\ref{fig:EffectTACMAC_dp50um_xO26_Tp} shows the effect on the temperature profile and $Z_\mathrm{O}$ for a \textcolor{black}{$10$\textmu m} and $50$\textmu m particle burning in $X_\mathrm{O_2} = 0.21$ and with $T_\mathrm{g} = 300 \: \mathrm{K}$. Both $\alpha_\mathrm{T}$ and $\alpha_\mathrm{m}$ have a significant effect on the particle temperature. \textcolor{black}{For the $10$\textmu m particle, the maximum temperature could vary by $200 \: \mathrm{K}$ due to the variations of $\alpha_\mathrm{T}$ and $\alpha_\mathrm{m}$, while for the $50$\textmu m particle the variation is around $130 \: \mathrm{K}$. The effect is more significant for the smaller particles, since they are more dominated by the free-molecular heat and mass transfer}.

An increasing $\alpha_\mathrm{T}$ results in a lower particle temperature, while the opposite is seen for a decreasing $\alpha_\mathrm{T}$. With an increasing $\alpha_\mathrm{T}$, the exchange of kinetic energy between the gas molecules and the surface is more efficient, resulting in a greater rate of heat loss and therefore a decreasing particle temperature. A variation in $\alpha_\mathrm{T}$ hardly affects the oxidation rate.

A variation in $\alpha_\mathrm{m}$ affects both the temperature profile as well as the oxidation stage $Z_\mathrm{O}$. An increasing $\alpha_\mathrm{m}$ results in a faster oxidation rate for the iron particle, leading to a faster heat release and therefore an increase in particle temperature. Note that the variation is relative, and thus the absolute difference with respect to the original values becomes smaller when $\alpha_\mathrm{m}$ is close to zero.

\subsection{Effect of transition modeling on the burn time}

\begin{figure}[h]
    \centering
    {\includegraphics[width=\figwsingle\columnwidth, clip]{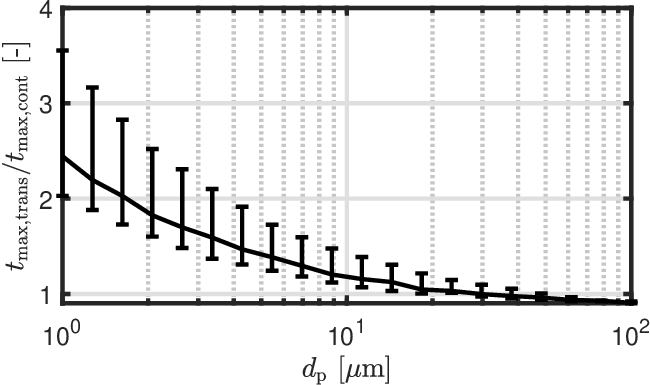}}
    \caption{The effect of the MD-informed Knudsen model compared to the continuum model for $t_\mathrm{max}/t_\mathrm{max,co}$. Values are obtained with $X_\mathrm{O_2}  = 0.21$.}%
     \label{fig:Basetimemax_tmaxco_vsdp_O21}
\end{figure}

The effect of the transition regime on the temperature profile of a variety of particles burning in $X_\mathrm{O_2} = 0.21$ and with $T_\mathrm{g} = 300 \: \mathrm{K}$ is investigated. Figure \ref{fig:Basetimemax_tmaxco_vsdp_O21} shows the difference for the time to maximum temperature $t_\mathrm{max}$ obtained with either the continuum and transition model as a function of particle diameter. \textcolor{black}{Only the oxidation to $Z_\mathrm{O} = 0.5$ is considered for the MD-informed Knudsen model. Here,  $t_\mathrm{max}$ is the time elapsed from $T_{\mathrm{p,0}} = 1100\:  \mathrm{K}$ to $T_{\mathrm{max}}$}. Due to the large difference in values for $t_\mathrm{max}$, the relative difference with respect to the continuum model is plotted. The error bars show the effect of varying both $\alpha_\mathrm{T}$ and $\alpha_\mathrm{m}$ by 50 \%.

When using the MD-informed Knudsen model, it is clear that the time to maximum temperature increases for smaller particles. Since the free-molecular regime inhibits mass transfer towards the particle, $t_\mathrm{max}$ increases. However, with an increasing particle size, the new $t_\mathrm{max}$ value becomes smaller than with the continuum model. This result is recognized as an effect of the decreasing MAC with an increasing $Z_\mathrm{O}$. A critical diameter $d_\mathrm{p,c}$ is defined, which means that if $d_\mathrm{p} < d_\mathrm{p,c}$, it is important to include the transition regime. For $d_\mathrm{p,c}$ we define that if one considers $t_\mathrm{max}/t_\mathrm{max,co} < 1.10$ as the criterion, critical particle size is found to be $d_\mathrm{p,c} \approx 15$\textmu m. This criterion suggests that, if one uses a continuum-based model to describe the combustion of iron particles smaller than $10$~\textmu m, the burn time can be underestimated by more than $10\%$ due to neglecting the effects of transition-regime transport phenomena.

\subsection{Effect of particle size on the maximum temperature}
\begin{figure}[h]
    \centering
    {\includegraphics[width=\figwsingle\columnwidth, clip]{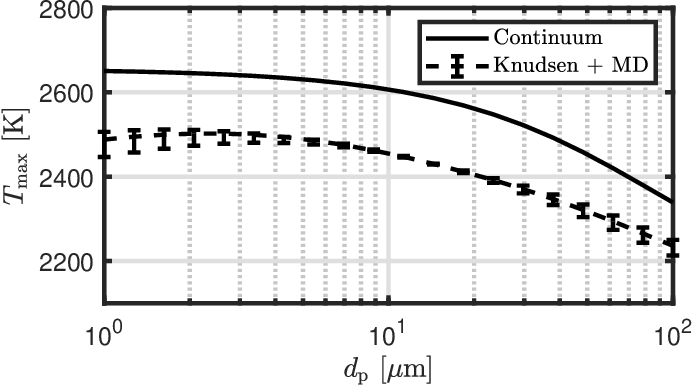}}
    \caption{The effect of the MD-informed Knudsen model compared to the continuum model for $T_\mathrm{max}$. Values are obtained with $X_\mathrm{O_2}  = 0.21$.}%
     \label{fig:BaseTmax_Tmaxco_vsdp_O21}
\end{figure}

Figure \ref{fig:BaseTmax_Tmaxco_vsdp_O21} shows the maximum temperature ($T_\mathrm{max}$) as a function of the particle size, obtained with either the MD-informed Knudsen model or the continuum model. The conditions are $X_\mathrm{O_2} = 0.21$ and $T_\mathrm{g} = 300 \: \mathrm{K}$. $T_\mathrm{max}$ as a function of particle size changes with respect to the continuum model. It is hard to distinguish any transition regime effects in this curve since due to the decreasing MAC with increasing $Z_\mathrm{O}$, the maximum temperature already decreased with respect to the continuum model. \cite{Ning2021} observed a decreasing maximum particle temperature with a decreasing particle size in the range of $25$-$54$\textmu m. \textcolor{black}{\cite{Panahi2022} did not observe the effect of particle size on the maximum temperature in the particle size range of $38$-$53$\textmu m. Judging from the PSD (particle size distribution) used by \cite{Panahi2022}, namely $38$-$45$\textmu m and $45$-$49$\textmu m, the particle samples did not differ significantly in diameter. To ensure that these distributions are not overlapping and thereby altering the influence of the particle size on the maximum temperature, a more comprehensive description of the PSD's used in the studies would be required.}
\textcolor{black}{In the present work, an even opposite trend is observed: The maximum temperature decreases with increasing particle size. Since both the amount of potential heat release and thermal inertia scale with $d_\mathrm{p}^3$, the $d_\mathrm{p}$ vs $T_\mathrm{max}$ trend can be predicted and explained by examining how the heat release rate $\dot{Q}_r$ and heat loss rate $\dot{Q}_l$ scale with the particle size $d_\mathrm{p}$ (denoted as $\dot{Q}_i\left(d_\mathrm{p}\right)$). If the heat loss rate increased faster with particle size than the heat release rate, then the maximum temperature would decrease with an increasing particle size. If the heat release rate increases faster with increasing particle size than the heat loss rate, and then the maximum temperature will increase. To obtain the decreasing maximum particle temperature with a decreasing particle size trend, the following inequality must hold
\begin{equation}\label{eq:inequality}
    \frac{\int_{t_0}^{t_{\mathrm{max}}}\dot{Q}_r\left(d_\mathrm{p}\right) \,dt}{\int_{t_0}^{t_{\mathrm{max}}}\dot{Q}_l\left(d_\mathrm{p}\right) \,dt} > 1.
\end{equation}}

\textcolor{black}{In completely external-diffusion-limited continuum models neglecting the effects of slip velocity and radiation, both the heat release rate and the heat loss rate scale with $d_\mathrm{p}$ (See Eq. \eqref{eq:mdot_cont} and Eq. \eqref{eq:qdot_cont}). Therefore, $\frac{\dot{Q}_r}{\dot{Q}_l} \propto \frac{d_\mathrm{p}}{d_\mathrm{p}} = 1$ and no particle size effect on the maximum temperature is expected. The slip velocity will affect how $T_\mathrm{max}$ scales with $d_\mathrm{p}$ since it affects both the Sherwood and Nusselt numbers, which have an impact on $\dot{Q}_r$ and $\dot{Q}_l$. Since for air the Lewis number is slightly larger than 1, it holds that $\mathrm{Sc} > \mathrm{Pr}$ and therefore $\mathrm{Sh}\left(d_\mathrm{p}\right) > \mathrm{Nu}\left(d_\mathrm{p}\right)$. As a result, the maximum temperature will increase with particle size. However, since the Lewis number is nearly one, the impact is negligible. \cite{Ning2021} argue that a larger particle has less radiation heat loss due to a smaller surface-to-volume ratio than a smaller particle. However, radiative heat loss scales with $d_\mathrm{p}^2$ and thus total heat loss scales with $d_\mathrm{p}^m$, with $m > 1$. So in the case of an external-diffusion-limited continuum model with radiation, $\frac{\dot{Q}_r}{\dot{Q}_l} \propto \frac{d_\mathrm{p}}{d_\mathrm{p}^m}$, which results in exactly the opposite trend as observed by \cite{Ning2021}: The maximum temperature decreases as particle size increases. This explains why the maximum temperature in the model decreases as particle size increases.}

\textcolor{black}{Due to the inclusion of chemisorption, the particle is not completely external-diffusion-limited. This results in a heat release rate that does not scale linearly in $d_\mathrm{p}$ (as it would for a fully external-diffusion-limited particle), but instead scales as $d_\mathrm{p}^n$, with $n > 1$. This effect could result in the inequality of Equation~\eqref{eq:inequality}. On the other hand, the impact of the free-molecular regime is a function of particle size. Therefore, this will also impact the heat release rate, so it holds
\begin{equation}
    \frac{\dot{Q}_r}{\dot{Q}_l} \propto \frac{d_\mathrm{p}^n}{d_\mathrm{p}^m}.
\end{equation}
The combined effect of chemisorption and transition-regime heat and mass transfer cannot be used to explain the lowering maximum particle temperature with a decreasing particle size trend.} 

\begin{figure}[h]
    \centering
    {\includegraphics[width=\figwsingle\columnwidth, clip]{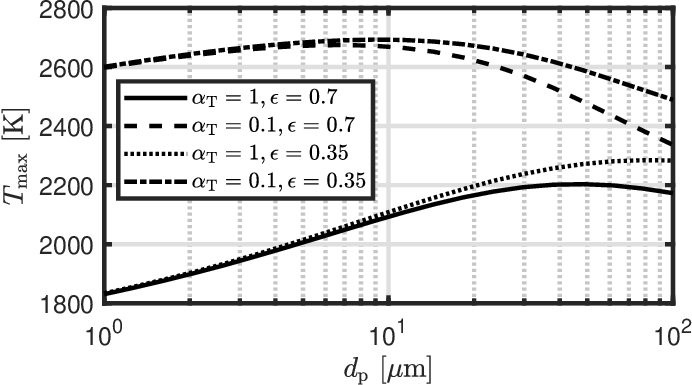}}
    \caption{The effect of the MD-informed Knudsen model compared to the continuum model for $T_\mathrm{max}$. Values are obtained with $X_\mathrm{O_2}  = 0.21$.}%
     \label{fig:Tmaxvsdp_effectTAC_eps}
\end{figure}

\textcolor{black}{To investigate the effect of particle size on the maximum temperature, a parametric study has been conducted. The maximum temperature vs particle size trend is affected by radiation and heat transfer in the Knudsen regime, as was previously stated. The TAC in this work is based on molecular dynamics simulations and is a function of $Z_\mathrm{O}$, with $epsilon = 0.7$ based on iron-oxide. (L2)\citep{Muller2015}. Figure \ref{fig:Tmaxvsdp_effectTAC_eps} demonstrates a sensitivity analysis of the TAC and emissivity. The emissivity is adjusted into the emissivity of liquid iron (L1) $epsilon = 0.35$ \cite{Muller2015}, while the TAC is varied between $\alpha_\mathrm{T} = 1$, and $\alpha_\mathrm{T} = 0.1$. Only the region above $10$\textmu m is impacted by the changing emissivity because radiative heat loss only becomes substantial for larger particles. Variations in emissivity between $0.7$ and $0.35$ can lower the maximum temperature for the large particle of $100$\textmu m by about $150 \: \mathrm{K}$. Small particles are affected significantly by changes in the thermal accommodation coefficient, with variations in the maximum temperature of about $800 \: \mathrm{K}$. Assuming the emissivity of liquid iron (L1) and assuming perfect thermal accommodation ($\alpha_\mathrm{T} = 1$) will result in an increasing maximum temperature with increasing particle size trend. However, perfect accommodation for metal-gas interactions is not expected. Therefore, further research is required to explain and confirm the particle size effect on the maximum temperature. }

\section{Conclusions}\label{sec:conclusions}
Molecular dynamics simulations have been performed to investigate the thermal and mass accommodation coefficients for the combination of iron(-oxide) and air. The obtained relations for the TAC and MAC are used in a point-particle Knudsen model to investigate the effects on the combustion behavior of (fine) iron particles.

The TAC for the interaction of $\mathrm{Fe}$ with $\mathrm{N_2}$ is almost independent of the surface temperature and equals $\alpha_\mathrm{T} = 0.17$. For $\mathrm{Fe_xO_y}$-$\mathrm{O_2}$ interactions, the TAC remains close to unity when the oxidation degree of the surface is low, but decreases abruptly to $0.2$ once it reaches the stoichiometry of $\mathrm{FeO}$. \textcolor{black}{The MAC decreases as a function of $Z_\mathrm{O}$. A fast decrease is observed if $Z_\mathrm{O} < 0.5$ and a weaker decrease if $Z_\mathrm{O} > 0.5$. This trend indicates that once the particle reaches stoichiometric $\mathrm{FeO}$, it becomes more difficult to absorb oxygen.}

By incorporating the MD information into the single iron particle model, a new temperature-time curve for the single iron particles is observed compared to results obtained with previously developed continuum models. Since the rate of oxidation slows down as the MAC decreases with an increasing oxidation stage, the rate of heat release decreases when reaching the maximum temperature, such that the rate of heat loss exceeds that of heat release, \textcolor{black}{while the particle is not yet oxidized to $Z_\mathrm{O} = 0.5$}. In addition, the oxidation beyond $Z_\mathrm{O} = 0.5$ (from stoichiometric $\mathrm{FeO}$ to $\mathrm{Fe_3O_4}$) is modeled. The effect of the transition-regime heat and mass transfer on the burn time becomes more than 10\% if the particles are smaller than $15$\textmu m.

In the cases with relatively high $\mathrm{O_2}$ concentrations in the gas-phase, the model overestimates the particle temperature. The reactive cooling slope observed after the maximum particle temperature, however, reasonably agrees with the experimentally observed slope. This overestimation could be attributed to either the assumption of an infinitely fast transport of oxygen inside the particle or to the validity point-particle model itself. These are areas that should be further investigated in future work.


The results and findings of the present work provide new insights into the effects of the free-molecular regime and surface chemisorption on the oxidation of single iron particles that could open up more comprehensive models for fine iron particle oxidation.

\section{Acknowledgements}\label{sec:Acknowledgements}
The authors acknowledge J.G.M. Kuerten and T. Hazenberg for useful discussions in developing this paper. We would like to thank D. Ning and A. Panahi for sharing the experimental data. We would like to thank B. Cuenot for initiating the collaboration between Eindhoven University of Technology and Imperial College London.

This project has received funding from the European Research Council (ERC) under the European Union’s Horizon 2020 research and innovation programme under Grant Agreement no. 884916. and Opzuid (Stimulus/European Regional Development Fund) Grant agreement No. PROJ-02594.

\appendix




\bibliography{biblio}







\end{document}